\documentclass[%
 reprint,
 amsmath,amssymb,
 aps,
superscriptaddress,
nofootinbib]{revtex4-2}

\usepackage{graphicx}
\usepackage{dcolumn}
\usepackage{bm}
\usepackage{gensymb}
\usepackage{wrapfig}
\usepackage{natbib}
\usepackage{float}
\usepackage{titlesec}
\usepackage{comment}
\usepackage{xcolor}
\usepackage{makecell}

\begin{document}

\title{Low-Energy Radon Backgrounds from Electrode Grids in Dual-Phase Xenon TPCs}

% LZ

% 1 
\author{D.S.~Akerib}
\affiliation{SLAC National Accelerator Laboratory, Menlo Park, CA 94025-7015, USA}
\affiliation{Kavli Institute for Particle Astrophysics and Cosmology, Stanford University, Stanford, CA  94305-4085 USA}

% 2 
\author{A.K.~Al Musalhi}
\affiliation{University College London (UCL), Department of Physics and Astronomy, London WC1E 6BT, UK}

% 3 
\author{F.~Alder}
\affiliation{University College London (UCL), Department of Physics and Astronomy, London WC1E 6BT, UK}

% 4 
\author{B.J.~Almquist}
\affiliation{Brown University, Department of Physics, Providence, RI 02912-9037, USA}

% LUX finished
\author{S.~Alsum}  
\affiliation{University of Wisconsin-Madison, Department of Physics, Madison, WI 53706-1390, USA}

% 5 
\author{C.S.~Amarasinghe}
\affiliation{University of California, Santa Barbara, Department of Physics, Santa Barbara, CA 93106-9530, USA}

% 6 
\author{A.~Ames}
\affiliation{SLAC National Accelerator Laboratory, Menlo Park, CA 94025-7015, USA}
\affiliation{Kavli Institute for Particle Astrophysics and Cosmology, Stanford University, Stanford, CA  94305-4085 USA}

% 7 
\author{T.J.~Anderson}
\affiliation{SLAC National Accelerator Laboratory, Menlo Park, CA 94025-7015, USA}
\affiliation{Kavli Institute for Particle Astrophysics and Cosmology, Stanford University, Stanford, CA  94305-4085 USA}

% 8 
\author{N.~Angelides}
\affiliation{Imperial College London, Physics Department, Blackett Laboratory, London SW7 2AZ, UK}

% 9 
\author{H.M.~Ara\'{u}jo}
\affiliation{Imperial College London, Physics Department, Blackett Laboratory, London SW7 2AZ, UK}

% 10 
% ...

%LUX

\collaboration{The LZ Collaboration}

\collaboration{The LUX Collaboration}

% LUX

\begin{abstract}
The dual-phase xenon time projection chamber (TPC) is a powerful technology to detect rare interactions such as scatters of dark matter particles on nuclei. In particular, the built-in gain of ionization signals in a dual-phase TPC makes it sensitive to events in the few-electron regime, as expected from low-mass dark matter interactions. The pursuit of this low-energy sensitivity through ionization-only signal detection has so far been hindered by excessive electron backgrounds observed across experiments.  
Much of this background is attributed to the plate-out of \(^{222}\)Rn decay chain isotopes on the high voltage electrode grid surfaces that span the full cross section of the TPC. This work presents a first-principle model constructed for this background, the predictions of which are consistent with data from the LZ and LUX experiments. We then discuss mitigation strategies of this background in future dual-phase TPCs and the possibility of applying this grid background model to ionization-only dark matter searches. 
\end{abstract}

\maketitle

\setlength{\parskip}{0pt}
\titlespacing{\subsubsection} {0pt}{6ex}{2ex}
\titlespacing{\subsection} {0pt}{6ex}{2ex}

\section{Introduction}
\label{sec:intro}

The dual-phase xenon time projection chamber (TPC) is a detector technology that is widely used for WIMP dark matter searches in low-background environments~\cite{LZFirstResults,XENONnT2025_FewEDM,XENON100_477days,LUXFirstResults}. The TPC is comprised of a target liquid xenon (LXe) volume bounded above by a thin layer of gaseous xenon (GXe).  
Energy depositions in the liquid produce a burst of prompt scintillation light (S1) and a number of freed atomic electrons. Static electric fields created by a set of high voltage electrode grids drift these electrons to the liquid surface and extract them into the gas, where the electrons are accelerated and produce delayed electroluminescence light (S2) proportional to the number of extracted electrons. S1 and S2 photons are typically observed using arrays of photomultiplier tubes (PMTs) at the top and bottom of the chamber. Together, the horizontal (\textit{xy}) position provided by the S2 light distribution in the PMTs and the vertical (\textit{z}) position provided by the time delay between the S1 and S2 enable a full 3-D reconstruction of an interaction location within the TPC, which allows for the definition of an inner fiducial volume where backgrounds are suppressed~\cite{AprileAndDokeReview}. 

LUX and LZ (LUX-ZEPLIN) are two dual-phase xenon TPC experiments that have operated within the last decade \cite{LUXFirstResults,LZDetector}. LUX was operated in the Davis Cavern at the Sanford Underground Research Facility in Lead, South Dakota, USA, and took WIMP search data between 2013 and 2016. After LUX was decommissioned in 2016, LZ was installed in the same facility in 2019 and began datataking in 2021. The LUX and LZ experiments have respectively reached spin-independent (SI) WIMP-nucleon cross sections of 1.1x10\(^{-46}\)cm\(^{2}\) (50 GeV/c\(^{2}\)) and 2.2x10\(^{-48}\)cm\(^{2}\) (40 GeV/c\(^{2}\)) with their standard WIMP search analyses \cite{LUXFullExposure,LZ2025_WS}.

Typical searches for WIMP dark matter in dual-phase xenon TPCs rely on observing both the S1 and S2 signals created by an energy deposition, but these searches rapidly lose sensitivity for energy depositions of $\mathcal{O}$(1)~keV or below due to non-observation of the S1. This is a result of the $\mathcal{O}$(10\(\%\)) efficiency for observing S1 photons. The probability of observing an ionization electron, however, is around 50-80\(\%\) in LUX and LZ, and values close to 100\% have been demonstrated in some detectors~\cite{PIXeY2018_EEE, Xu2019_XeEEE}. As a result, many low-energy recoils without S1s can still be observed in the S2 channel. In this energy regime, one can regain sensitivity to low-mass dark matter scatters using an ``S2-only'' analysis, in which the requirement on the S1 is removed~\cite{Essig2012_SubGeVXENON10,XENON1TS2O,KelseyLUX,XENONnT2025_FewEDM}. This lower-threshold analysis provides sensitivity to dark matter candidates with masses much lower than those probed by S1+S2 searches, down into the MeV/c\(^{2}\) regime in some models. By considering the Migdal effect, in which nuclear recoils may be accompanied by electron recoils~\cite{Migdal}, S2-only searches in dual-phase xenon TPCs can further improve their low-mass dark matter sensitivity~\cite{XENON1T2019:Migdal,SubGeVLUX}. These searches can also improve the statistics in measurements of the coherent elastic neutrino-nucleus scattering (CE\(\nu\)NS) of solar \(^{8}\)B neutrinos on xenon, which is a physics signal of interest in its own right and in the context of xenon-based WIMP search experiments~\cite{LZResult8B2025,B8,PandaX2025_8B,XENONnT2025_8B}.

While an S2-only analysis is a powerful tool for increasing a xenon TPC's sensitivity at low energies, that power comes at the cost of increased backgrounds. Notably, the lack of an S1 and corresponding drift time makes rejection of backgrounds from the top and bottom of the detector challenging. Moreover, a number of additional pathological electron backgrounds are particularly impactful in S2-only searches, among which are spurious field-induced emission~\cite{ImperialWireStudies,LZ2025_EBg}, radiogenic emission from high voltage grids~\cite{Linehan}, and delayed electrons following large S2s~\cite{LUXEBackgrounds,LZ2025_EBg}. 
Together, these additional backgrounds contribute to a steep rise in the S2 rate in the few-electron regime as observed by recent experiments, limiting the discovery power of S2-only searches~\cite{XENON1TS2O,KelseyLUX}.

One of these challenging S2-only backgrounds comes from surface decays of \(^{222}\)Rn daughters on the stainless steel (SS) wires of the cathode and gate electrode grids. 
These radon daughters primarily originate from plate-out on the wires during the grid fabrication process \cite{Linehan,EricDRIFT,OtherDrift,ThirdDrift}, though some may be deposited in-situ from radon emanated into a TPC during operation. This plate-out background is expected to be dominant over intrinsic radioimpurities in the wires, and the strong and nonuniform electric fields near the grid wires can suppress detectable S1 and S2 signals, making these decays a prominent contributor to S2-only backgrounds in these detectors. As a result, a dedicated effort to characterize these backgrounds is valuable.

This work  
develops a first-principles model to predict the S1 and S2 responses of this radon background on grid surfaces in dual-phase xenon TPCs. In Section \ref{sec:Physics}, we discuss the origins and properties of the relevant radionuclides that give rise to this background. We follow with Section \ref{sec:ModelConstruction}, in which we present the construction of a model for this background, predicting the TPC's S1 and S2 responses to radon progeny decays on high-voltage grid wires. 
Section \ref{sec:S1S2Models} compares the model predictions with data from LZ and LUX. Section~\ref{sec:Discussion} discusses how this background may be mitigated in future experiments, presents a quantitative extension of this model into the S2-only regime, and comments on the role of this background in shaping the S2-only spectrum in dual-phase xenon TPC experiments. 
We conclude in Section \ref{sec:Conclusions}.

\section{Radon-Induced Grid Backgrounds}
\label{sec:Physics}

Radon is a chemically inert radioactive gas produced in the decay chains of trace uranium (U) and thorium (Th) contaminants found in many materials. Of all radon isotopes, \(^{222}\)Rn has a half-life that is long enough (3.8 days) to enable its diffusion through the material in which it is produced and subsequent migration into lab and detector spaces.\footnote{\(^{220}\)Rn has a short half life of 55.6~s, which limits its emanation from detector components; LZ measured bulk \(^{220}\)Rn chain activities 30--100 times lower than those of \(^{222}\)Rn and its progeny~\cite{LZBackgrounds}.} The decay products of \(^{222}\)Rn are not chemically inert and may plate out on a surface, quasi-permanently affixing themselves to that surface. Subsequent decays can then occur over decades, creating problematic backgrounds. 
In this section, we discuss the origins of \(^{222}\)Rn daughter plate-out on a TPC's high-voltage grids, and which of the decays in the \(^{222}\)Rn chain are the most relevant as backgrounds in low-energy dark matter searches.

\begin{figure}[t!]
\centering
\includegraphics[width=\linewidth]{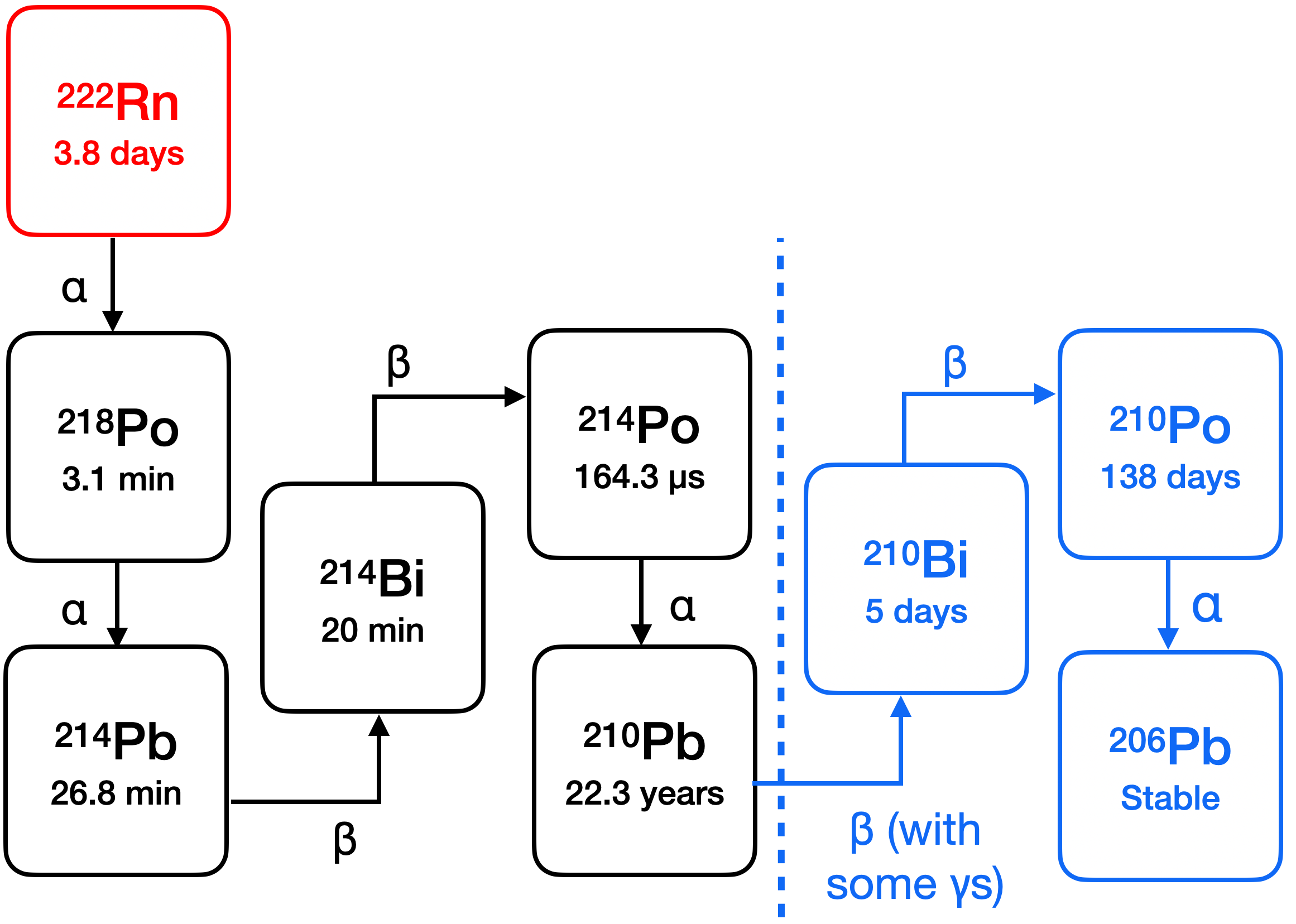}
\caption{The decay chain for \(^{222}\)Rn, with half-lives and decay modes. We have highlighted in blue the set of decays that are expected to be observed during TPC operation due to \(^{210}\)Pb plateout on the wires during grid production, and have left in black the set of decays that may be observed due to in-situ \(^{222}\)Rn emanation during TPC operation (see discussion in Sec.~\ref{subsec:SourcesOfRadonChainPlateouBackgrounds}).}
\label{fig:RadonChain}
\end{figure}

\subsection{Origins of \(^{222}\)Rn-chain Plate-out Backgrounds}
\label{subsec:SourcesOfRadonChainPlateouBackgrounds}

There are two channels through which \(^{222}\)Rn daughters may plate out on grid wires. In the first, daughters can be deposited on the wire surfaces during grid production and installation when \(^{222}\)Rn decays in ambient air near the grid. Though concentrations of \(^{222}\)Rn in ambient air vary with geographic location, U/Th contamination of nearby objects, and deliberate mitigation strategies for reducing ambient radon, typical concentrations at the surface of the earth are around 10 Bq/m\(^{3}\)~\cite{WHO}. Following the decay of a \(^{222}\)Rn nucleus, decays from its daughters \(^{218}\)Po, \(^{214}\)Pb, \(^{214}\)Bi, and \(^{214}\)Po proceed within approximately an hour (Figure~\ref{fig:RadonChain}), leaving \(^{210}\)Pb, which has a 22.3 year half-life. Because of this long half-life, decays of \(^{210}\)Pb and its daughters, \(^{210}\)Bi and \(^{210}\)Po, can be observed long after radon gas exposure ends, such as during TPC operation. 

The ratios of the measured decay rates of \(^{210}\)Pb, \(^{210}\)Bi, and \(^{210}\)Po are dependent on the time between the grid's last exposure to \(^{222}\)Rn-rich air and the measurement. Secular equilibrium is only approximately achieved if this period is long enough to let the \(^{210}\)Po decay rate ``grow-in.'' As the half-life of \(^{210}\)Po is 138~days, this occurs approximately one year after the last grid exposure to air. 
For LUX and LZ, this condition was easily met because the datataking did not begin until $\sim$1 and $\sim$2 years, respectively, after the grids were installed and the detectors were sealed~\cite{ManninoThesis,LinehanThesis}. %\footnote{
While radon emanation from detector components during these ``waiting'' periods may contribute radon exposure, the resulting plate-out is estimated to be subdominant (Appendix~\ref{app:AdditionalModelDetails}).

\begin{table*}[htpb]
\centering
\setlength\extrarowheight{2pt}
\caption{Approximate characteristic ranges of various decay products present in the source decays of concern. Decay products of non-alpha decays occur according to their branching ratios, and are not necessarily equal in rate. Continuous-slowing-down-approximation (CSDA) ranges are quoted for electrons, and are overestimates of the expected range. Attenuation lengths are given for x-rays and gammas. Data are from \cite{NISTXRays,ESTAR,SRIM}. As the decay of \(^{214}\)Bi is rather complicated with many deexcitation gammas, we do not list specific ones but recognize that there are a variety of possible gamma and beta energy combinations from this decay. The same is true for \(^{210}\)Pb, in which many of the decay products involve conversion electrons, Auger electrons, and characteristic X-rays that are not given dedicated table entries.}
\begin{tabular}{|w{c}{2.5cm}||w{c}{2.7cm}|w{c}{2.7cm}|w{c}{3.1cm}|w{c}{2.7cm}|}
\hline
Parent Nuclide & Decay product & Energy & Range (LXe) & Range (SS304) \\
& & [keV] & [\(\mu\)m] & [\(\mu\)m] \\
\hline
\hline
 \(^{218}\)Po & \(\alpha\) & 6000 & 49.8 & 12.4 \\
 & \(^{214}\)Pb & 112 & 0.071 & 0.018 \\
\hline
 \(^{214}\)Pb & \(\beta\) & $<$1018 & $<$2500 & $<$770 \\
\hline
 \(^{214}\)Bi & \(\beta\)'s + \(\gamma\)'s & \(<\)3269 (endpoint) & Varied & Varied \\
\hline
\(^{214}\)Po & \(\alpha\) & 7686 & 66.8 & 17.8 \\
  & \(^{210}\)Pb & 146 & 0.0807 & 0.021 \\
\hline
 \(^{210}\)Pb & \(\beta\) & $<$17 & $<$4.9 & $<$1.4 \\
  & \(\beta\) & $<$63.5 & $<$38.2 & $<$11.6 \\
  & X-ray & 15.7 & 66 & 25 \\
  & \(\gamma\) & 46.5 & 216 & 532 \\
\hline
 \(^{210}\)Bi & \(\beta\) & $<$1161.5 & $<$2700 & $<$900 \\
\hline
 \(^{210}\)Po & \(\alpha\) & 5304 & 41.9 & 10 \\
 & \(^{206}\)Pb & 103 & 0.066 & 0.017 \\
\hline
\end{tabular}
\label{table:RelevantDecayProducts}
\end{table*}
\vspace{5mm}

Some fraction of the \(^{210}\)Pb plateout may be embedded in the top layer of wire material by alpha decays of \(^{218}\)Po and/or \(^{214}\)Po. If a \(^{218}\)Po atom lands on a wire surface and decays with the alpha ejected away from the wire, the daughter \(^{214}\)Pb recoils with sufficient energy to embed itself approximately 20~nm (and up to \(\sim 40\)~nm) into a SS304 wire, with a final depth dependent on the angle of the alpha with respect to the surface normal~\cite{SRIM}. The later \(^{214}\)Po decay enables the \(^{210}\)Pb to further embed itself in the same fashion. Since alpha decays are isotropic, the depth profile of \(^{210}\)Pb follows a relatively smooth distribution that falls off around 20-25~nm from the wire surface. If the \(^{214}\)Pb, \(^{214}\)Bi, or \(^{214}\)Po lands on the wire, then only one such ``embedding'' process may occur, resulting in a distribution with a smaller characteristic embedding depth. 
Without direct knowledge of
which \(^{222}\)Rn daughters landed on the wire, it is  difficult to produce an a-priori distribution of the depth of the remaining \(^{210}\)Pb.
But as is to be discussed in Section~\ref{sec:S1S2Models}, an ``embedding fraction," \(f_{e}\), defined as the fraction of on-wire decays that occur below the surface, can be obtained by comparing model prediction to observed data. 

The second channel through which \(^{222}\)Rn daughters can plate out onto the grids is during TPC operation: radon emanated in-situ by TPC components can yield daughters that migrate to and plate out onto the grid wires. This migration can happen either under the influence of xenon fluid flow or, if the daughters are electrostatically charged, as a result of electrostatic drifts along field lines in the detector~\cite{EXO200IonMobility,MallingThesis,BradleyThesis,LZRadonFlow2025,McLaughlinThesis}. A notable difference between this channel and the ambient-air plate-out channel is that  
the early \(^{222}\)Rn chain daughters \(^{218}\)Po, \(^{214}\)Pb, \(^{214}\)Bi, and \(^{214}\)Po that plate-out onto the wires  
can also produce a background in the TPC. These early-chain decays will therefore be important to consider while building a comprehensive low-energy model of plated-out \(^{222}\)Rn-chain backgrounds. Despite the importance of modeling these short-lived daughters, the rate of subsequent \(^{210}\)Pb-chain decays from in situ emanation is highly suppressed by the long half-life of \(^{210}\)Pb (see Appendix~\ref{app:AdditionalModelDetails} for more discussion).
The overall background contribution from emanation is also suppressed by the low \(^{222}\)Rn emanation rates achieved in xenon TPCs~\cite{BradleyPaper,XENONnTRadonRemoval,ChottLZEmanation}.

\subsection{\(^{222}\)Rn Chain Decays Relevant for Low-Energy Searches}
\label{subsubsec:RelevantDecays}

Table~\ref{table:RelevantDecayProducts} summarizes information about the decay radiation from \(^{222}\)Rn daughters. Decay products can be broadly separated into two classes that are useful  
for understanding grid backgrounds. The first class is a set of ``high-energy,'' several-MeV alphas from \(^{218}\)Po, \(^{214}\)Po, and \(^{210}\)Po, which are easily recognizeable in a TPC due to their monoenergetic spectra. With these large energies, these events can be isolated in analysis and provide a measurement of the decay rates of the underlying radionuclides. Due to the isotropic nature of alpha decays, only approximately 50\(\%\) of decays near the wire surface deposit the full alpha energy into xenon, and hence the activity of the radionuclide is approximately double the measured alpha rate.\footnote{For cylindrical wires, a small fraction of alphas may ``clip through'' the wire and re-enter the xenon~\cite{EricDRIFT}. However, for LUX and LZ, the wire diameters are large enough that no more than $\sim$1.5\(\%\) of events are expected to fall under this category.} 

The second class of relevant decay products is a set of ``low-energy'' (\(<70~\)keV\(_{ee}\)) radiation which defines the shape of the S2 spectrum from grid backgrounds in both the low-energy S1+S2 region and the S2-only dark matter search region. This set includes \(^{210}\)Pb decays, the low-energy components of the \(^{210}\)Bi and \(^{214}\)Pb spectra, and the recoiling \(^{206}\)Pb (\(^{214}\)Pb) nucleus from \(^{210}\)Po (\(^{218}\)Po) decays where the alpha is lost in the wire. While a small fraction of \(^{214}\)Bi decays will be in this energy range, the rapidly-following \(\alpha\)-decay from \(^{214}\)Po makes it easy to identify these events and remove them from low-energy searches for new physics. The rest of these low-energy decay products are the core components of the low-energy models we produce in this work.

\section{Grid Model Construction}
\label{sec:ModelConstruction}

Understanding the S1 and S2 signals observed during decays from radon-induced grid background relies on knowing properties of both the background's origin and of the detector response. In this section we describe the construction of a quasi-detector-agnostic model of radiogenic grid backgrounds in a dual-phase xenon TPC. 

Even though plate-out backgrounds are expected to be present on all grids in a dual-phase TPC, we will construct this model only for the gate and cathode, as these grids are the only ones with event topologies relevant to an S2-only search (See Appendix~\ref{app:AdditionalModelDetails} for more discussion). Our model is built on four sub-components: the spatial distribution of \(^{222}\)Rn daughters and resulting decay radiation, the light and charge yields in extreme electric fields, the near-wire light collection efficiency, and the electron transport from grid wire surfaces. While the materials and electric fields referenced in this and following sections are LUX- and LZ-specific, the method in which this model is constructed can be readily generalized to other TPCs assuming comparable field conditions.

\subsection{Spatial Distribution of \(^{222}\)Rn Daughters and Decay Radiation}
\label{sec:RnSpatial}

\begin{figure}[t!]
\centering
\includegraphics[width=\linewidth]{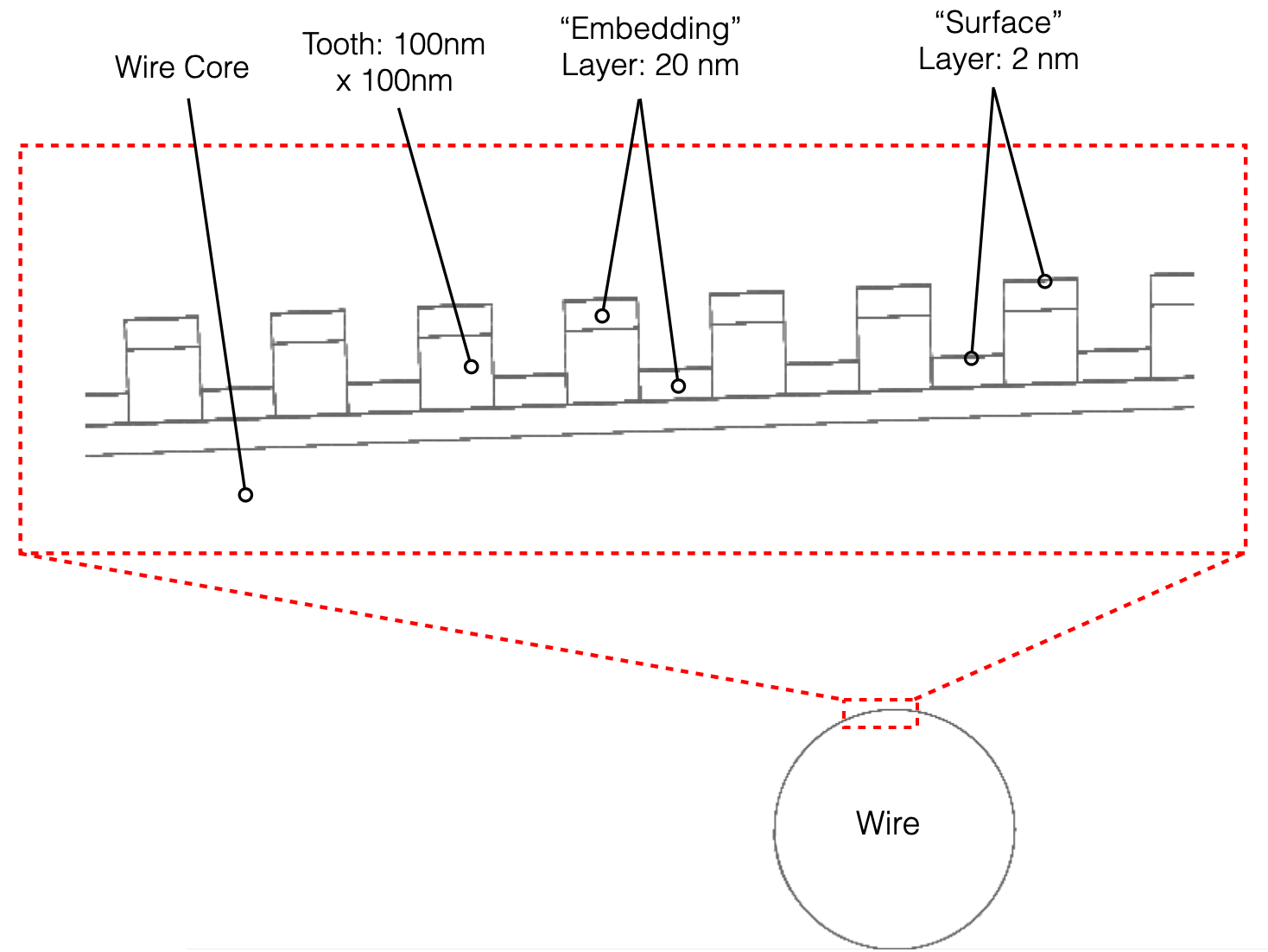}
\caption{A close-up of the BACCARAT dedicated geometry used for simulating events. Teeth are added to model wire surface roughness, and both ``embedded'' and ``surface'' layers of the geometry are included. An inner layer used to generate embedded decays in a no-teeth, smooth-wire geometry is also shown, but is not used to generate any of the models explicitly presented in this work.}
\label{fig:RoughWireGeometry}
\end{figure}

To model decays from the relevant \(^{222}\)Rn daughters, we use the Geant4-based BACCARAT simulation framework~\cite{BACCARAT} with an additional inclusion of Geant4's StandardNR physics list for capturing subtle energy degradation processes in alpha decays due to ion-ion scattering within wire materials. We build a dedicated SS304 crossed-wire-cell-in-LXe simulation geometry, and simulate decays uniformly within either a 2-nm-deep ``surface'' layer or a 20-nm-deep ``embedded layer'' of SS (Figure~\ref{fig:RoughWireGeometry}). A non-uniform depth profile, as predicted with SRIM simulations~\cite{SRIM,LinehanThesis}, is only expected to marginally degrade the modeled \(^{206}\)Pb recoil energy peak in this work. An additional set of longitudinal striations, or ``teeth,'' are included along the length of the wire to capture subtle degradation of decay energies by wire surface roughness. These teeth are modeled as having a 100~nm \(\times\) 100~nm profile, with the 100~nm spacing informed by scanning-electron microscope (SEM) measurements of grid wire used for prototyping LZ's grids and the 100~nm depth assumed to match the spacing. The teeth also inherit the surface and embedded layers described above.
This surface roughness primarily affects sub-micron electron tracks with keV-level total energy. While the \(^{206}\)Pb and \(^{214}\)Pb ion recoil responses are also affected, ion energy loss in a tooth is partially compensated by alpha energy deposition in liquid xenon between tooth structures, in the direction opposite to the recoiling ion's momentum. Different scales of tooth depth and width were studied and yielded only marginal ($\sim20\%$) variation in the event rates observed from this baseline model in the S2 region between [0,20]e- that is relevant for S2-only searches, though all tooth models deviated noticeably from the smooth-wire model. Above about 30e-, only marginal differences are observed between the S2 spectra of all wire profile models studied. 

\begin{table*}[t!]
\centering
\setlength\extrarowheight{2pt}
\caption{Computed values of the saturation region for the LUX and LZ gate and cathode. The saturation region (``Sat. Reg.'') values depend on the voltages and geometry of the wires considered. For LUX values, see Ref~\cite{ManninoThesis}. For LZ values, see Ref~\cite{LinehanThesis}.}
\begin{tabular}{|w{c}{3.4cm}||w{c}{2.0cm}|w{c}{2.4cm}|w{c}{3.2cm}|w{c}{5.8cm}|}
\hline
Grid & Wire Radius  & Surface Field & Sat. Reg. Radius & Sat. Reg. Distance from Wire Surface \\
& [\(\mu\)m] & [kV/cm] & [\(\mu\)m] & [\(\mu\)m] \\
\hline
\hline
LUX Gate (Run3) & 51 & 45.3 & 226.5 & 175.5 \\
LUX Cathode (Run3) & 103 & 16.5 &  165 & 62 \\
LZ Gate (WS2022) & 37.5 & 36.5 & 137 & 99.5 \\
LZ Cathode (WS2022) & 50 & 16.3 & 81 & 31 \\
\hline
\end{tabular}
\label{table:SaturationRegionValues}
\end{table*}
\vspace{5mm}

For each \(^{210}\)Pb-chain component, we simulate both surface and embedded populations to provide an extra degree of modeling flexibility in lieu of knowing the distribution of daughters' embedding depths from ambient-air plate-out during grid production. \(^{218}\)Po and \(^{214}\)Pb are simulated as having only a ``surface'' component resulting from internal \(^{222}\)Rn emanation. While it is technically possible for \(^{214}\)Pb to become embedded from decays of \(^{218}\)Po that land on grid wires, the short half-life of \(^{218}\)Po implies that most emanated \(^{218}\)Po ions would have decayed before drifting to the grids, so  
the majority of the \(^{214}\)Pb found on the wire will have landed there as \(^{214}\)Pb. 
Even for detectors where convective xenon fluid flow is fast and governs ion transport (causing a large fraction of \(^{218}\)Po to plate-out before decaying), a ``surface-only'' \(^{214}\)Pb  model is still a good approximation because the large, MeV-scale energy of the \(^{214}\)Pb \(\beta-\)decay will only be minimally impacted by the $\sim$20 nm embedding. Decays of \(^{214}\)Bi and \(^{214}\)Po are not simulated in our model because their close proximity in time makes them overwhelmingly appear together, enabling efficient removal of these backgrounds from data. A \(^{214}\)Bi-\(^{214}\)Po decay pair may evade the coincidence tagging if 
the \(^{214}\)Bi has an extremely low-energy beta decay unaccompanied by gammas \textit{and} the subsequent \(^{214}\)Po has the alpha lost into the wire with an ion recoiling into liquid xenon. However, these pathological pairs 
are expected in no more than a few percent of all \(^{214}\)Bi-\(^{214}\)Po decays on the wire. The spatial distribution of daughters in azimuth around each wire's axis is dependent on the isotope, grid history, grid position in the TPC, and TPC operating conditions, assumptions which are discussed in later detector- and grid-specific sections of this work.

\subsection{Light and Charge Yields}
\label{subsubsec:LightAndChargeYields}

The partitioning of deposited energy between production of S1 photons and freed electrons near a grid wire is more complicated than that for interactions in the bulk xenon, far from electrode surfaces. This is in large part due to the field variation: near smooth cylindrical grid wires, the electric field rises as 1/\(r\) with decreasing distance \(r\) to the wire center, culminating in wire surface fields at the 10~kV/cm scale, much higher than the \(\mathcal{O}(100\)~V/cm) fields found in the TPC bulk. Within these regions of increased electric field, charges freed in a near-wire energy deposition have a smaller chance to recombine with xenon ions before they are pulled away from the interaction site, leading to higher charge (Q) yield and lower light (L) yield relative to the bulk. As the field strength falls off with increasing distance from the wire, these Q (L) yields will  
gradually decrease (increase) to their values in the bulk liquid over several wire radii from the wire surface. 

However, this yield variation is nonlinear in field strength: while the Q/L yields for betas vary by a factor of a few between $\sim$100~V/cm and 5~kV/cm, yields show significantly reduced field variation above this field strength range due to a ``bottoming-out'' of the electron-ion recombination at the event site~\cite{NEST,Dahl,Aprile1991}, a pattern also observed for nuclear recoils~\cite{LLNLStudy2019}. As a result, there is a cylindrical ``shell'' of xenon around each wire in which, even though the field changes rapidly, its value is large enough (\(\gtrsim 10\)~kV/cm) that the Q/L yields are mostly constant. 
We call this shell the ``yield saturation region,'' or YSR. For TPCs which can hold sufficiently high voltage to make the YSR thickness \(\gtrsim 30~\mu\)m, a majority of the \(^{222}\)Rn daughter decay products relevant to low-energy searches will have tracks fully contained within the YSR (Table~\ref{table:RelevantDecayProducts}). This coincidence is the foundation on which an approximate model of these grid backgrounds can be built: it allows one to approximate Q/L yields for these backgrounds with a single ``high-field'' value for each species, rather than performing a complicated analysis of spatially-varying yields.

The LUX and LZ gate and cathode grids largely satisfy this requirement (Table~\ref{table:SaturationRegionValues}).
We also restrict ourselves to considering only events with deposited energy \(<\)150~keV,
which further ensures that modeled events overwhelmingly occur within the YSR. Further study of events \(>\)150~keV, such as most of the decays of \(^{210}\)Bi, could yield marginal gains in understanding these radon plate-out backgrounds, but only at the cost of a significant increase in modeling difficulty due to varying Q/L yields. For events below this 150~keV ``energy cutoff'', we query Q/Y yields corresponding to the surface field values in Table~\ref{table:SaturationRegionValues}  
from the NEST package (version 2.2.1) to compute the number of photons and electrons generated\footnote{As an example, yields at these fields for \(\beta\)-decays are approximately 58 e-/keV and 12 photons/keV at 60~keV. For ion recoils, yields are handled with an event energy clustering scheme that also takes into account the ion-dependent yields of Fe, Ni, and Cr ions from the stainless steel wire. For more detail we refer the reader to Ref.~\cite{LinehanThesis}.}~\cite{NEST}. With this approximate model we anticipate
capturing the leading-order dependence of the grid S2 spectra on the electric field.

\begin{figure}[t!]
\centering
\includegraphics[width=0.82\linewidth]
{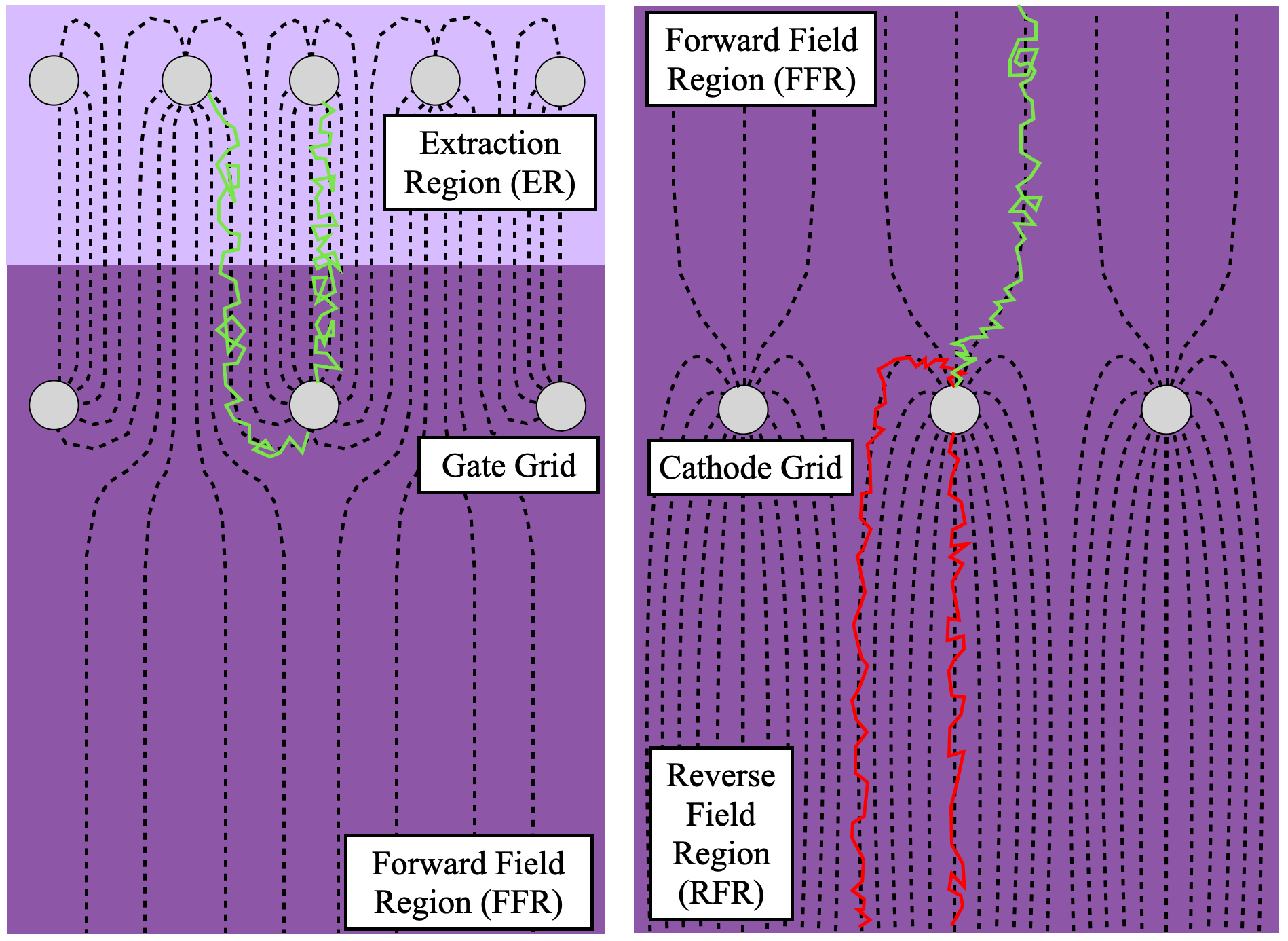}
\includegraphics[width=\linewidth]
{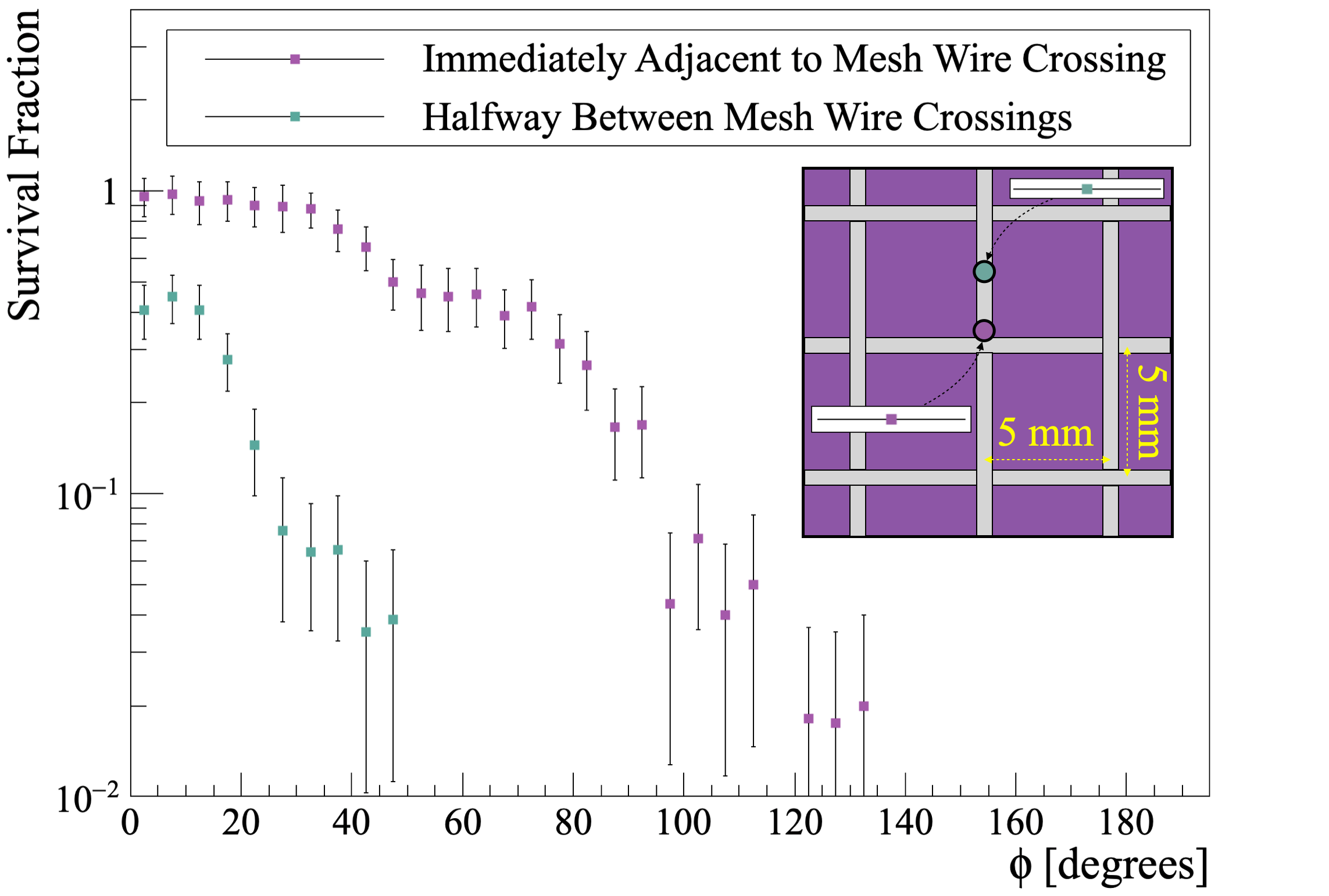}
\caption{\textbf{Top:} a side-view diagram illustrating how the field line structure around the gate (left) and cathode (right) impacts tracks of electrons starting at the grid wires. Green tracks indicate electrons that drift toward the extraction region, and red tracks are electrons lost to the RFR. \textbf{Bottom:} simulation-based estimates of the charge survival fractions for the LZ cathode, as a function of azimuthal location \(\phi\) around the wire circumference. Here, 0\(\degree\) implies the top of the wire, and 180\(\degree\) implies the bottom of the wire. Purple (green) points show the survival fraction for a location laterally very close to (far from) a wire crossing on the LZ cathode, with lateral positions indicated in the accompanying top-down diagram. These are computed using Garfield++~\cite{Garfieldpp} with high-field diffusion parameters estimated from Ref~\cite{Boyle}.}
\label{fig:CombinedFFRSurvivalFig}
\end{figure} 

\subsection{Light Collection}
\label{subsubsec:LightCollectionEfficiency}

\begin{table*}[t!]
\centering
\setlength\extrarowheight{2pt}
\caption{Parameters used in building models for the LZ and LUX gate and cathode background models. Brackets indicate parameter ranges that define the bounds of a grid's model family. Some parameters (such as electron lifetime) which were not constant over the course of the datataking period were described by singular representative values for that time period. In rows three and four, ``azimuthal distribution'' refers to the isotope distribution around the wire circumference.}
\begin{tabular}{|w{c}{7.2cm}||w{c}{5.8cm}|w{c}{3.4cm}|}
\hline
Model Parameter & Model Values (LZ) & Model Values (LUX) \\
\hline
\hline
Source radioisotopes (Gate) & \(^{210}\)Pb chain & \(^{210}\)Pb chain \\
Source radioisotopes (Cathode) & \(^{210}\)Pb chain, \(^{218}\)Po, \(^{214}\)Pb &  \(^{210}\)Pb chain \\
\(^{210}\)Pb-chain azimuthal distribution  & \makecell{Uniform, Uniform + 10\(\%\) excess spread \\ over top 10\(\degree\), 40\(\degree\), 120\(\degree\)} & Uniform \\
\(^{218}\)Po and \(^{214}\)Pb azimuthual distribution (cathode) & Uniform over top [18\(\degree\),36\(\degree\)] & N/A \\
Maximum energy for model applicability & 150~keV\(_{ee}\) & 150~keV\(_{ee}\) \\
Cathode S1 cutoff & 158~phd & 158~phd \\
\textit{g1} (Gate) & [0.077,0.081] & [0.069,0.073] \\
\textit{g1} (Cathode) & [0.1,0.122] & [0.093,0.109] \\
Extraction Efficiency & [0.781,0.849] & [0.46,0.52] \\
Electron Lifetime & 5500\(\mu\)s & 750\(\mu\)s \\
\hline

\end{tabular}
\label{table:BaseModelParametersTable}
\end{table*}
\vspace{5mm}

In the near-wire region, there is an additional reduction in the observed S1 due to a lowered S1 light collection efficiency (LCE) relative to that in the xenon bulk. As S1 light is emitted isotropically, approximately 50\(\%\) of the light from an energy deposition located at the wire's surface will initially interact with the SS wire, whose relatively low reflectivity \((<57\%~\cite{ssReflectivity})\) will cause a non-negligible reduction in observed light. This LCE also depends on the energy deposition's radial and azimuthal location around the wire and the grid's \(z\) distance to the two PMT arrays. To capture this LCE position dependence, we made dedicated optical simulations for each grid in the full LZ geometry. For each grid, we simulated photons a few microns above the wire top (0\(\degree\) azimuth) and tabulated a bound of that grid's LCE using the photons reaching the PMT arrays. We performed the same treatment with photons launched from the wire underside (180\(\degree\) azimuth) to find another bound to the LCE, which, taken with the top-side simulation, gives the \(g_{1}\) range presented in Table~\ref{table:BaseModelParametersTable}. For LZ, optical properties of the TPC followed the values outlined in the optimistic model defined in Ref. \cite{tdr}, but with a SS-LXe reflectivity of 0.45, derived from an analysis comparing S1 areas of cathode and near-cathode alphas. We estimated LCE values for LUX by scaling the LZ values using the ratio of the overall detector g1 values between LZ and LUX.

For each event, this computed LCE is used to determine the number of detected S1 photons from the set of photons created in the Q/L modeling step.

\subsection{Electron Transport and S2 Response}
\label{subsec:NearWireFieldStructure}

Once freed charge has been produced, the near-wire field structure governs the fraction of that charge that drifts up to the liquid surface and is observed. Since the extraction field is typically much stronger than the drift field in a TPC, all field lines landing on the gate originate from the anode, so electrons from gate wire events are expected to drift upward into the extraction region to produce an S2, even for those produced on the lower part of the wires (Figure~\ref{fig:CombinedFFRSurvivalFig}). Charge collection from field-shadowed regions of a grid, such as where wires touch in a woven-wire mesh, may be reduced due to the lower charge yield and a possibility of electron re-absorption into the wire surface. However, the relative contribution of these is expected to be low so a quantitative assessment is not included in this work.
The S2 spectrum from decays on the gate is thus expected to have an overall shape similar to that of the underlying energy spectrum.

The cathode field structure is qualitatively different: in LUX and LZ, only approximately 10\(\%\) of field lines ending on the cathode originate from the anode, the other 90\(\%\) being from the bottom grid or bottom PMT array. This 90-10 ratio is determined by the  
$\sim$10 times higher field strength in the reversed field region (RFR) than that in the forward field region (FFR). 
For similar experiments where the magnitude of the cathode voltage is much higher than any other voltage in the TPC, this ratio is fundamentally driven by the relative heights of the FFR and RFR. This behavior begins to break down if the cathode voltage magnitude is low enough that the gate or bottom shield voltages play a non-negligible role in determining the drift or RFR fields, respectively.

The cathode's near-wire field structure has a profound effect on the S2 response for near-cathode events. As illustrated in Figure \ref{fig:CombinedFFRSurvivalFig}, a drifting electron created at the wire surface will diffuse laterally and longitudinally. 
If the electron diffuses onto field lines that lead into the extraction region, then it may survive to contribute to an S2. The S2 response for a full electron cloud near the cathode can be constructed by treating each electron in the cloud independently and summing the electrons that survive~\cite{Boyle}. 
The probability of electron detection depends on the initial azimuthal position of the electron around the wire: near the top of the wire this ``survival fraction'' is high, while near the bottom of the wire it is vanishingly small (Figure \ref{fig:CombinedFFRSurvivalFig}, bottom). For crossed-wire or woven-wire meshes (like in LZ), this survival fraction also changes substantially depending on the distance from a wire crossing.

Because of this response, even very energetic, multi-MeV energy depositions may contribute just a handful of electrons if they occur near the underside of the wire. This somewhat nonintuitive charge loss effect causes the S2 values to be highly smeared to very low values, resulting in very weak correlation between deposited energy and S2 size. As we will show, these RFR losses give the cathode S2 spectrum a distinctly different shape than that from the gate. As LZ and LUX had geometrically different cathode structures, a dedicated FFR survival function was computed for use with each.

Once we account for this near-wire field response in tandem with the spatial distribution of decays around the wire, we model the rest of the S2 response. This includes capturing the electron lifetime losses for the grid-to-liquid-surface drift time, the extraction efficiency, and a parametrized SE electroluminescence gain. After these responses, the model returns an S2 area in photons detected (phd), which is then normalized using the detector SE area to produce a number of electrons, \(Ne\)-. A summary of model parameters, including some discussed in more detail in the following section, is given in Table~\ref{table:BaseModelParametersTable}.

\section{Model Comparison with LZ and LUX Data}
\label{sec:S1S2Models}

We now produce detector-specific implementations of the generic model described in Section~\ref{sec:ModelConstruction}, and compare these model predictions to low-energy gate and cathode S1+S2 spectra measured by the LZ and LUX experiments during datataking runs in 2022 (WS2022) and 2013 (Run3), respectively. These low-energy spectra were produced using the data analysis procedure and cuts described in Appendices~\ref{app:SelectingLowEnergyDatasetsLZ} and~\ref{app:SelectingLowEnergyDatasetsLUX}.

\begin{table*}[t!]
\centering
\setlength\extrarowheight{2pt}
\caption{Measured alpha rates from the LZ gate and cathode wires. The ``Implied Decay Rate'' column is a factor of two higher than the ``Measured Alpha Rate'' column -- it compensates for the fact that 50\(\%\) of the \(\alpha\)-particles go into the wire and don't appear in the measured \(\alpha\) rate. These total rates are those used for normalization of the low-energy models. The rightmost column labeled ``Missing TPC Rate" is constructed using bulk alpha measurements from Ref.~\cite{LZBackgrounds}. It represents the difference between the overall TPC \(^{222}\)Rn rate and the corresponding TPC daughter alpha rate. For cathode \(^{214}\)Po and \(^{218}\)Po, we confirm that this ``missing'' rate is comparable with our on-cathode measured values in the ``Implied Decay Rate'' column. See text for futher discussion.}
\begin{tabular}{|w{c}{2.9cm}||w{c}{3.3cm}|w{c}{3.2cm}|w{c}{3.8cm}|w{c}{3.2cm}|}
\hline
Alpha & Measured Alpha Rate & Method & Implied Decay Rate & ``Missing'' TPC Rate \\
& [mHz] & & [mBq] & [mBq] \\
\hline
\hline
\(^{210}\)Po (Gate) & 5.6 \(\pm\) 0.3 & S1+S2 measurement & 11.2 \(\pm\) 0.6 & N/A\\
\(^{210}\)Po (Cathode) & 4.4 \(\pm\) 0.4 & S1-only measurement & 8.8 \(\pm\) 0.8 & N/A \\
\(^{214}\)Po (Cathode) & 7.8 \(\pm\) 0.9 & S1-only measurement & 15.7 \(\pm\) 1.8 & 14.9 \(\pm\) 2.7\\
\(^{218}\)Po (Cathode) & 1.1 \(\pm\) 0.2 & S1-only measurement & 2.2 \(\pm\) 0.4 & -0.28 \(\pm\) 3.3\\
\hline

\end{tabular}
\label{table:AlphaModelParametersTable}
\end{table*}
\vspace{5mm}

\subsection{Grid Radon Backgrounds in LZ}
\label{subsec:LZData}

The general prescription we use to produce grid  S2 background spectra for LZ's gate and cathode grids is as follows. For each grid, we first identify alpha decays of on-wire \(^{222}\)Rn chain daughters and measure their rates, which are then used to anchor rates of non-alpha decays in the \(^{222}\)Rn chain. For each \(^{222}\)Rn daughter, we then assume a distribution describing its azimuthal locations around the grid wire, and use this with 
the model described in Section~\ref{sec:ModelConstruction} to 
predict its contribution to the low-energy S2 background spectrum. Here, we keep only events with S1\(\geq3\) phd for comparison to S1+S2 data with identifiable grid origins. 

For the LZ gate, a single prominent population of single-scatter events corresponding to the gate drift time was found with an S1 \(\sim27\times 10^{3}\)~phd, which we matched to \(^{210}\)Po alpha decays. We made an analysis cut in log(S1)-log(S2) space to isolate these events, including those where the alpha loses part of its energy below the wire surface. We then converted the measured rate of these alphas (Table~\ref{table:AlphaModelParametersTable}) into an implied decay rate, which was then used for all three \(^{210}\)Pb chain components in the model, given that secular equilibrium was reasonably reached for this chain.

\begin{figure}[t!]
\centering
\includegraphics[width=\linewidth]{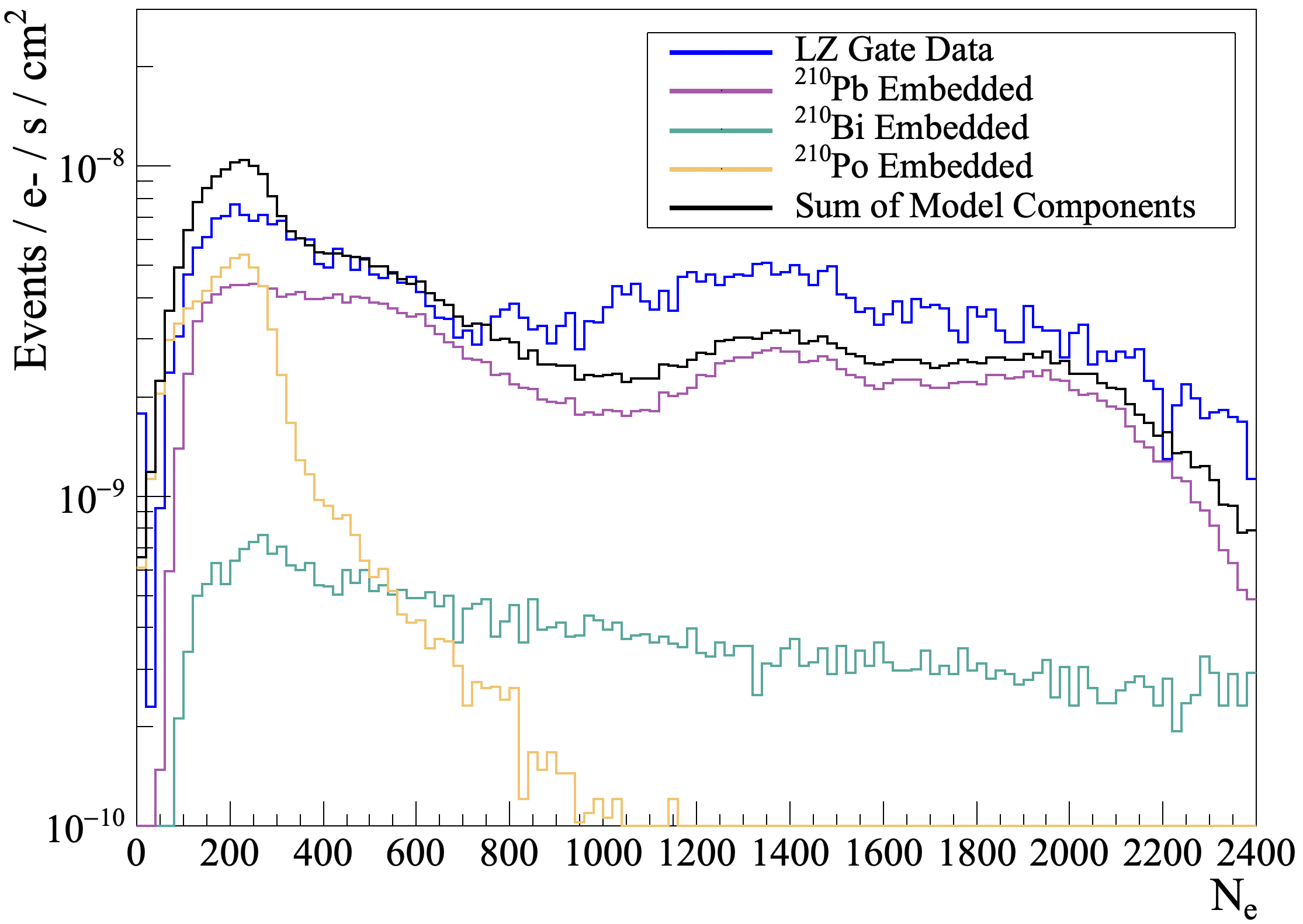}
\includegraphics[width=\linewidth]{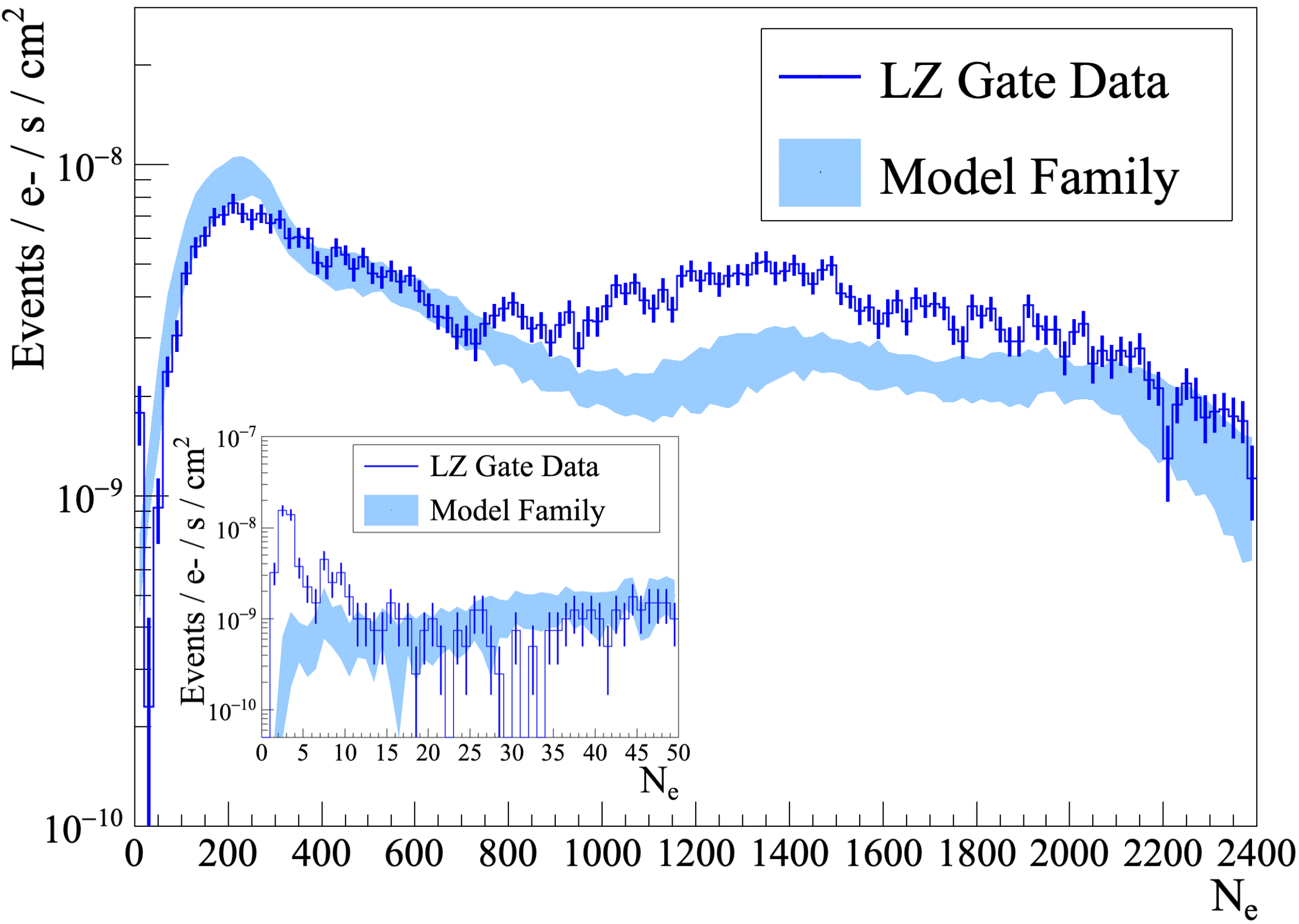}
\caption{\textbf{Top:} Data-model comparisons for example gate model, with component breakdowns. This model assumes an extraction efficiency of 0.78 and a near-gate \(g_{1}\) of 0.077, and embedding fraction \(f_{e}=1.0\). A normalization \(^{210}\)Po rate of 11.8~mBq (corresponding to the +1\(\sigma\) scenario in Table~\ref{table:AlphaModelParametersTable}) is used. The y-axis is normalized to the livetime and the total grid wire surface area, accounting for data quality cut acceptances. See Table~\ref{table:SaturationRegionValues} for wire geometry information and Fig. 7.23 from Ref.~\cite{LinehanThesis} for additional embedding scenarios. \textbf{Bottom:} comparison of the gate data with a family of models created by taking variations of model parameters in Table~\ref{table:BaseModelParametersTable}.} 
\label{fig:GateModelBreakdownPlusFamily}
\end{figure}

The lack of other prominent alpha populations on the LZ gate (beyond \(^{210}\)Po) suggests a minimal impact of in-situ plateout from emanated \(^{222}\)Rn daughters
on this grid, and that most plated-out \(^{210}\)Pb-chain daughters were from grid production. Our baseline model therefore assumes a uniform \(^{210}\)Po distribution in the azimuthal angle, since there is largely no preferred direction of plateout in ambient air. To account for the exposure of the grids to downward cleanroom air flow during fabrication, which could increase the plate-out rate on the top of the wires, we also consider scenarios in which there is a slight (10\(\%\)) excess of \(^{210}\)Pb-chain daughters distributed over the top 10\(\degree\), 40\(\degree\), or 120\(\degree\) of the wire.

We then use the generic model described in Section \ref{sec:ModelConstruction}, detector parameters listed in Table \ref{table:BaseModelParametersTable}, and the \(^{210}\)Po rates in Table~\ref{table:AlphaModelParametersTable} to produce a model of the gate's low-energy S2 spectrum. 
In this process, we adopt an embedding fraction parameter \(f_{e}\) (defined in Sec.~\ref{subsec:SourcesOfRadonChainPlateouBackgrounds}) to provide an additional degree of model flexibility given the lack of prior knowledge about the depth distribution of the \(^{210}\)Pb-chain ions from ambient-air plateout.

The top panel of Figure~\ref{fig:GateModelBreakdownPlusFamily} shows a component-by-component breakdown of an example model with \(f_{e}=1\) for the gate compared with data, for the S2 range of [0,2400]e-.
The peak in the spectrum around 250~e- is dominated by the monoenergetic \(^{206}\)Pb recoils from \(^{210}\)Po decays. 
If we instead assume all events originate on the surface of the wire (\(f_{e}=0\)), the \(^{206}\)Pb recoil peak would be 4--5 times higher and substantially exceed what is observed in data. 
We further produced a set of spectra using a range of \(f_{e}\) values and used a \(\chi^{2}\) fit with the measured gate S2 spectrum to identify a preferred \(f_{e}\). We found the best-fit \(f_{e}\) for the gate to be consistent with 1.0.

Away from the \(^{206}\)Pb peak, the model-data agreement is also reasonable for both the overall event rate and the spectral shape. 
A steep fall in the spectrum below 100~e- is attributed to the S1 threshold, which suppresses S1+S2 events at low energies. 
Both the model and data also drop off together above 2000~e-, near where the \(^{210}\)Pb decay endpoint (63.5 keV) is found. The data excess in the range of [750,2200]e- cannot be explained by \(^{210}\)Pb, the rate of which is strongly constrained by the observed \(^{210}\)Po alpha activity. 
Instead, we attribute this unmodelled background to \(^{125}\)I decays on the gate. This background is also observed in the LZ detector as a result of neutron activation from the DD calibration~\cite{TwoNeutrinoDECLZ2024}. Because the DD neutrons were aimed at the top of the TPC's liquid volume, \(^{125}\)I was likely produced near the gate grid, and due to iodine's halogenic properties, it is known to readily adsorb onto stainless steel surfaces~\cite{WrenIodine,BeckIodine}. \(^{125}\)I decays on the grids can deposit energies between 30~keV and 67~keV, as confirmed by an analysis in S1-S2 space that shows two populations in this energy region. Their rates decay with a half-life consistent with that of \(^{125}\)I (59 days), and the summed rate averaged over our datataking period is qualitatively consistent with the integrated discrepancy between the data and our model in the region between [750,2200]e-.

To evaluate systematic uncertainties associated with the data-driven parameters used by the model, we construct a family of models for the gate grid by generating all combinations of the parameters within their uncertainties as listed in Table \ref{table:BaseModelParametersTable}. In a given model, each parameter was toggled to either the top or bottom of its range, or in the case of three or more discrete options (such as with the azimuthal distribution), to each option. 
The alpha-based overall rate normalizations are also toggled between their minimum and maximum values as determined by the mean rates \(\pm1\sigma\). 
The minimum and maximum possible values in each S2 bin from the model family are plotted as a band in the lower panel of Figure~\ref{fig:GateModelBreakdownPlusFamily}, 
where the inset shows the low-energy part of the spectra. 
The discrepancy between the gate model and data below \(\simeq\)12e- is primarily attributed to  field-induced electron emission accompanied by photon production~\cite{ImperialWireStudies,Linehan,BaileyThesis,LZ2025_EBg}. Such grid emission is observed in LUX and LZ during otherwise stable TPC operations, and, differently from grid radiogenic backgrounds, the electrons emitted tend to be spatially and temporally clustered.

For the LZ cathode, rates of the various alpha populations were determined using an analysis (separate from that described in Appendix~\ref{app:SelectingLowEnergyDatasetsLZ}) in which only S1 information was used. This was done to avoid event detection inefficiencies due to the smeared S2 response from RFR electron losses (Section~\ref{subsec:NearWireFieldStructure}). Populations of \(^{218}\)Po, \(^{214}\)Po, and \(^{210}\)Po were observed on the cathode with the decay rates shown in Table \ref{table:AlphaModelParametersTable}. \(^{218}\)Po and \(^{214}\)Po decays on the cathode are attributed to in-situ plateout from emanated radon in the TPC. Charged \(^{222}\)Rn daughters in the TPC may drift downward in the applied drift field and reach the cathode because the convective flow in LZ is subdominant to the ion drift velocity in field (0.45~mm/s)~\cite{LZBackgrounds}.\footnote{\(^{218}\)Po and \(^{214}\)Po populations were not found in any appreciable amount on the gate because the liquid volume above the gate wires is between two and three orders of magnitude smaller than that above the cathode.} As a result, LZ observed a difference between the bulk TPC rates of \(^{222}\)Rn and \(^{218}\)Po (\(^{214}\)Po)~\cite{LZBackgrounds}. This ``missing'' TPC component of each daughter isotope is accounted for by the on-cathode rate of that isotope, as shown in Table~\ref{table:AlphaModelParametersTable}. For \(^{218}\)Po, the smallness of our measured on-cathode rate (2.2 mBq) relative to the total in-TPC \(^{222}\)Rn rate (33.4~mBq) is consistent with the picture that \(^{218}\)Po decays quickly in the liquid (Sec.~\ref{sec:RnSpatial}), and bulk-to-grid transfers are limited to a small vertical region extending only \(\mathcal{O}(8\)~cm) around the cathode. No on-grid populations associated with alpha decays in the \(^{220}\)Rn chain could be confidently observed above background, consistent with prior radioactivity studies in the LZ TPC~\cite{LZBackgrounds}.

\begin{figure}[t!]
\centering
\includegraphics[width=\linewidth]{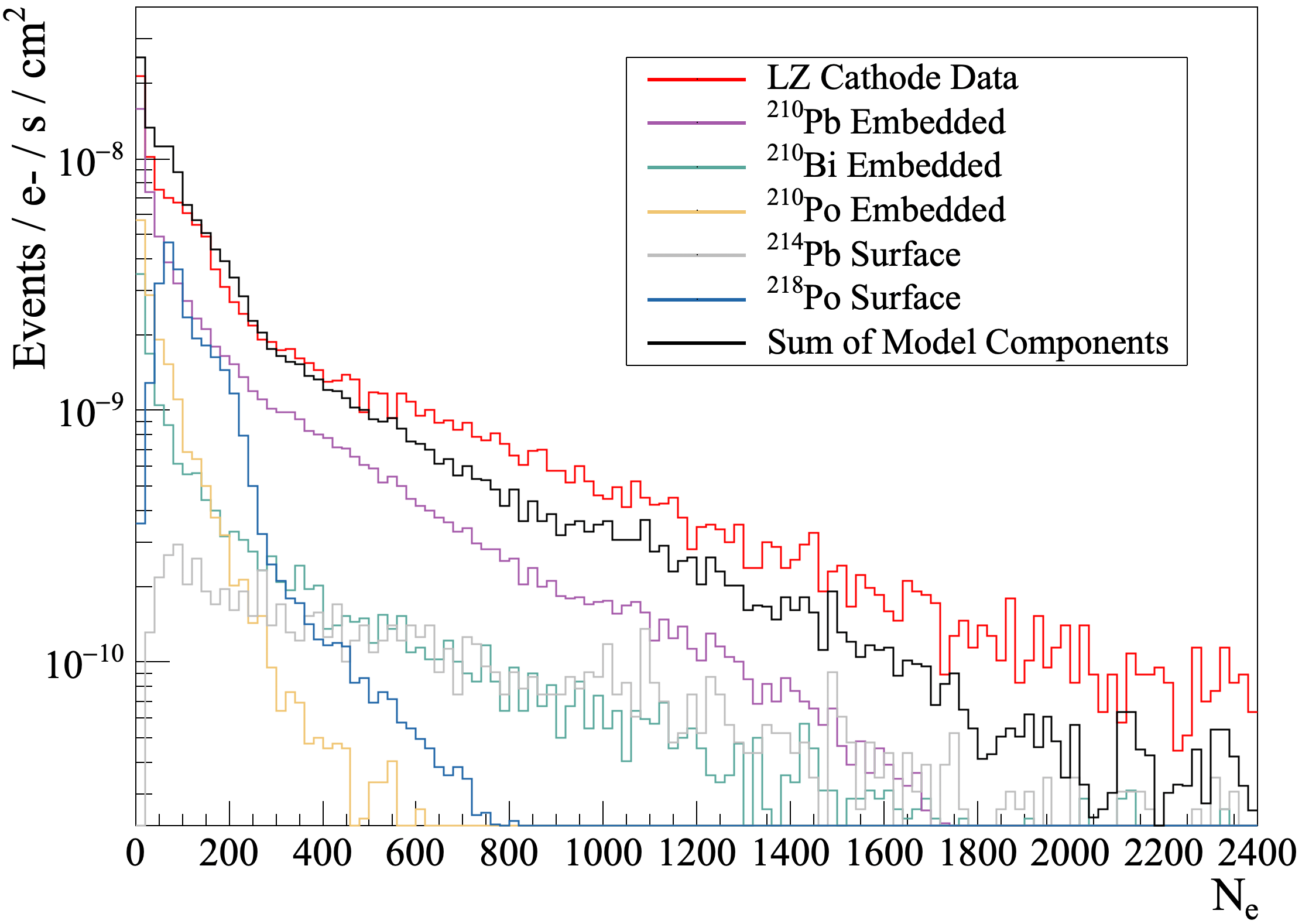}
\includegraphics[width=\linewidth]{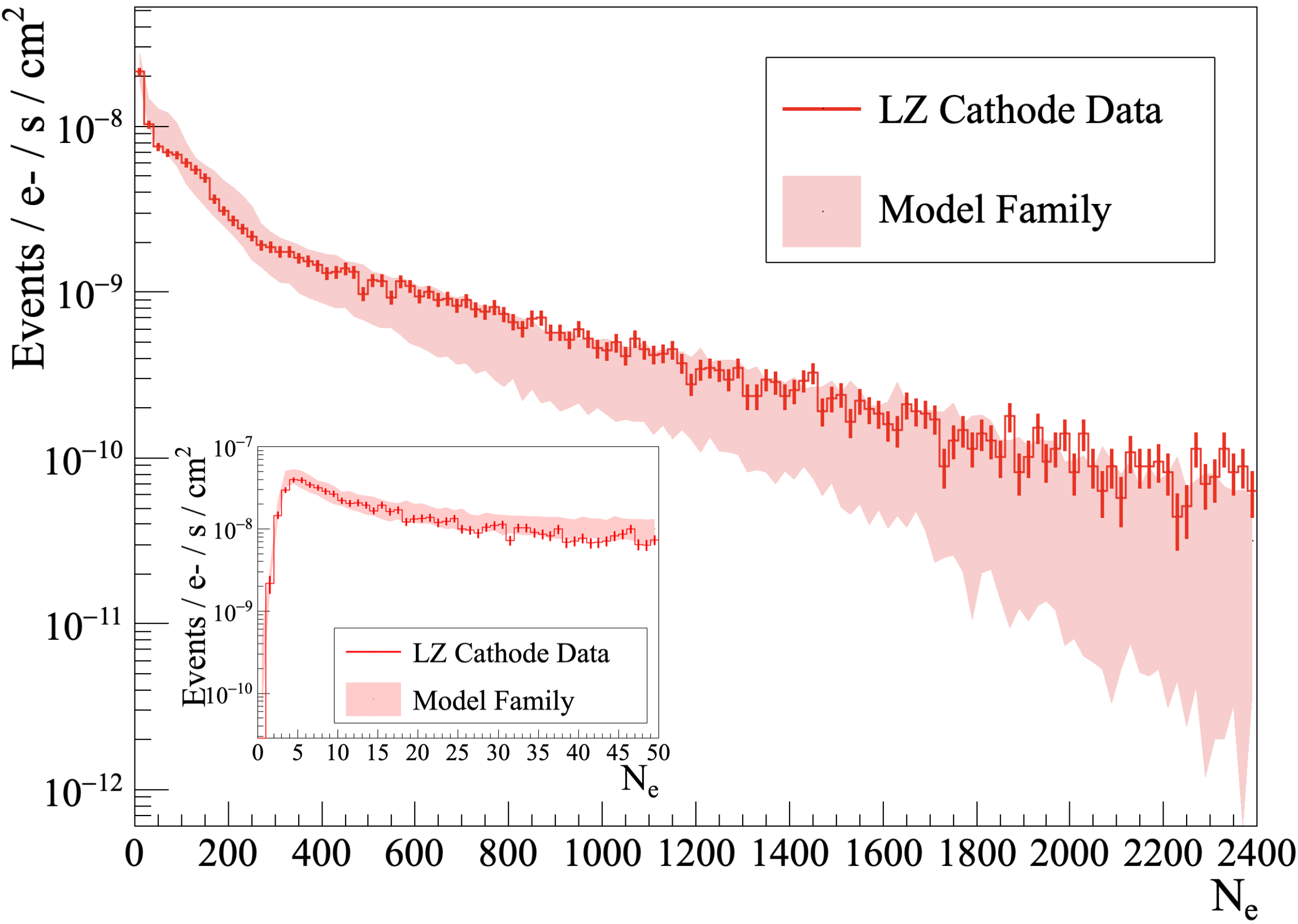}
\caption{\textbf{Top:} Data-model comparisons for an example cathode model, with component breakdowns. This model assumes an extraction efficiency of 0.78 and a near-cathode \(g_{1}\) of 0.1, and an embedding fraction \(f_{e}=1.0\) for the \(^{210}\)Pb-chain daughters. For the cathode, the normalization \(^{218}\)Po/\(^{214}\)Pb/\(^{210}\)Po rates are 2.6~mBq/17.5~mBq/9.6~mBq, and in-situ-deposited daughters are modeled as landing on the top 10\(\%\) of the wire with an \(f_{e}=0.0\). The y-axis is normalized to the livetime and the total grid wire surface area, accounting for data quality cut acceptances. \textbf{Bottom:} comparison of the cathode data with a family of models created by taking variations of model parameters in Table~\ref{table:BaseModelParametersTable}.}
\label{fig:CathodeModelBreakdownPlusFamily}
\end{figure}

While we expect the azimuthal distribution of ambient-air \(^{210}\)Pb plate-out to be the same for the cathode as for the gate, we expect the relevant \(^{218}\)Po and \(^{214}\)Pb populations to be ``top-heavy'' on the wire due to downward drift of charged daughters along FFR field lines. 
Based on the ratio of the FFR and RFR field strengths, the FFR-origin ions should land primarily on the top 10\(\%\) (36\(\degree\)) of the wire (Figure~\ref{fig:CombinedFFRSurvivalFig}, top right).  
We also consider a scenario where this estimate is tightened to the top 18\(\degree\) of the wire for comparison. While \(^{222}\)Rn decays in the RFR can contribute additional on-wire activities that populate the bottom of the cathode wires, their rates are suppressed due to the small volume of the RFR. More importantly, these RFR-origin surface decays primarily produce ionization signals that drift back to the RFR and will not be detected. We comment that the cathode \(^{218}\)Po and \(^{214}\)Pb rates reported above are obtained using cuts that prioritize above-cathode alpha decays because they are the most important for surface background modeling. Therefore, the model uncertainty from RFR \(^{222}\)Rn decays is expected to be relatively small.

We then use the techniques described in Section \ref{sec:ModelConstruction} and parameters listed in Table \ref{table:BaseModelParametersTable} to produce a model for the cathode S2 spectrum. The \(^{210}\)Pb-chain components are scaled by the measured \(^{210}\)Po alpha rates, and the \(^{218}\)Po decays (which give \(^{214}\)Pb recoils) are scaled by the measured \(^{218}\)Po alpha rate. The \(^{214}\)Pb decay rate was bracketed between those of on-grid \(^{218}\)Po and \(^{214}\)Po. 
The resulting cathode model, with contributions from different nuclides, is shown in the top panel of Figure \ref{fig:CathodeModelBreakdownPlusFamily}, compared to data selected using an S1-S2 analysis as described in Appendix~\ref{app:SelectingLowEnergyDatasetsLZ}. The global decline in event rate with increasing S2 size is a consequence of the charge loss to the RFR: high-energy events with large S1s can be found with S2s consisting of just a few electrons, which props up the lowest-S2 range of the spectrum.  In producing this spectrum, we remove contributions from highly-degraded high-energy events by invoking an ``S1 cutoff'' demanding S1\(<\)158~phd, which corresponds to an energy of roughly 83~keV\(_{ee}\). This S1 cutoff is more conservative than the 150~keV energy cutoff described in Section~\ref{subsubsec:LightAndChargeYields} to ensure that the YSR condition is met; at the same time, it allows the endpoint of \(^{210}\)Pb decays to be captured, and encompasses the vast majority of \(^{206}\)Pb and \(^{214}\)Pb recoils. 
In addition, events above this energy have long tracks, and thus are only minimally affected by the near-wire effects (Appendix~\ref{app:AdditionalModelDetails}).
The same S1 cutoff is applied in the analysis of LZ data for a consistent spectral comparison.

The shoulder-like feature around 150~e- is consistent with a population of monoenergetic decays primarily on the top of the wire, where smearing from charge loss into the RFR is minimal. We attribute this shoulder to surface \(^{214}\)Pb recoils at the top surface of the wire from the decays of downward-drifted \(^{218}\)Po (blue curve) following radon emanation in the TPC. 
For the \(^{210}\)Pb chain components, we produced several cathode spectra with varying embedding fractions \(f_{e}\), and through a \(\chi^{2}\) minimization process found it to be consistent with 1.0~\cite{LinehanThesis}. The systematic excess of the data over the model in the region above 450~e- may relate to our choice of approximating our woven wire mesh in COMSOL as a set of unwoven crossed wires. We expect this subtle difference to have a small impact on the average field structure, but may shift the charge survival fraction near the top of the wire (Figure~\ref{fig:CombinedFFRSurvivalFig}, bottom) where high-S2 events are produced. 
The over-prediction of events below 200~e- might indicate a less top-heavy spatial distribution of \(^{218}\)Po on the cathode wires due to RFR-origin radon activities. 
The bottom panel of Figure~\ref{fig:CathodeModelBreakdownPlusFamily} shows a model family constructed in the same way as that for the gate, and shows reasonable data-model agreement over a wide range of S2, including at the lowest S2 values modeled (the inset).

With both the LZ gate and cathode data-model comparisons complete, we now note a feature common to both: a near-unity embedding fraction \(f_{e}\) for the \(^{210}\)Pb-chain components. This makes natural sense for the gate, which went through a chemical passivation at the end of its production. This passivation removed $\sim$20\AA~of metal from the grid wire surfaces and may have also removed any surface component to \(^{210}\)Pb components already plated out. However, the cathode was not passivated, and still shows \(f_{e}=1\). This may be in part due to the fact that both the cathode and gate underwent a final pre-installation DI rinse: since some component of radon daughters is known to be attached to dust particulates that cannot burrow into a surface~\cite{Sun_2020,OTAHAL2025107781}, any such component would likely be removed with a rinse. Beyond this effect, we also hypothesize that the embedding fraction for \(^{210}\)Pb-chain components plated out onto fine wire grids during production in a cleanroom with HEPA-filtered airflow is naturally very close to one. \(^{222}\)Rn daughters (but not \(^{222}\)Rn itself) on the upstream side of a HEPA filter will likely plate out in the filter, ensuring that only \(^{222}\)Rn enters the cleanroom through the HEPA filter. For \(^{210}\)Pb to land on the wire surface, in-air decays from \(^{222}\)Rn to \(^{210}\)Pb would have to occur. As this characteristic timescale, \(\mathcal{O}(40\)~min), is much longer than the time for laminar airflow to push the \(^{222}\)Rn past the wires, it is most likely that if a radon daughter lands on the wires, it is one between \(^{218}\)Po and \(^{214}\)Bi, and not \(^{210}\)Pb~\cite{LinehanThesis}. This implies that at least one alpha decay will occur from the wire surface prior to \(^{210}\)Pb being deposited, ejecting ions either outward away from the wire or inward into the wire. Outward ejected ions have sufficient energies to travel many tens of microns from the wire in air. Since the wires are only \(O(100~\mu m)\) in diameter, these ejected ions may be swept away by cleanroom airflow and not redeposit onto the wires, leaving a naturally mostly-embedded distribution of \(^{210}\)Pb.

\subsection{Grid Radon Backgrounds in LUX}
\label{subsec:LUXData}

The general prescription to modeling the LUX gate and cathode backgrounds is similar to that for LZ. However, because of the reduced accessibility of LUX data and analysis tools, the rates of alpha decays on the grids cannot be easily obtained to normalize the S2 energy spectra. Instead, we fit the measured grid S2 spectrum in the low-energy region to our grid models to estimate the surface radioactivity levels. This approach is supported by the LZ study that confirms grid surface radioactivity as the dominant source of backgrounds near the grid region. In addition, the model developed for LZ is demonstrated to reproduce LZ data with high fidelity without significant tuning.

To reduce the number of components needed for the fits to the LUX grid S2 spectra, we also made a simplifying assumption that all grid decays can be approximated with \(^{210}\)Pb-chain components. This simplification was applied to both the fit components and the resulting model families constructed. Though this strategy ignores \(^{214}\)Pb and \(^{218}\)Po decays that were observed on the LZ cathode, the simplification can be justified for two reasons. First, due to the high endpoint energy of the \(^{214}\)Pb decay, its contribution to the low-S2 region is heavily suppressed relative to \(^{210}\)Pb for reasonable rates of \(^{222}\)Rn emanation;  in LZ, the \(^{214}\)Pb contribution only becomes significant past the \(^{210}\)Pb endpoint (Figure~\ref{fig:CathodeModelBreakdownPlusFamily}). 
Second, the primary spectral feature of \(^{218}\)Po is a recoil peak within a narrow energy range effectively identical to that from \(^{210}\)Po decays. Unlike \(^{214}\)Pb, this component does have a significant effect on fit quality, but it may also be effectively modeled just by varying the embedding fraction \(f_{e}\) for the \(^{210}\)Pb chain: a large amount of surface \(^{218}\)Po would enhance this ion recoil peak relative to the rest of the spectrum (\(f_{e}\simeq0\)), and negligible \(^{218}\)Po would lead to a less prominent peak (\(f_{e}\simeq1\)). As a result, a fit strategy where both \(f_{e}\) and the overall rate normalization of the \(^{210}\)Pb chain can float has the flexibility to absorb the absence of \(^{214}\)Pb and \(^{218}\)Po while still achieving a reasonable fit and estimating overall activities. Similar to the LZ work, \(^{210}\)Pb chain decays are treated as in secular equilibrium, and uniformly distributed over the wire surface.

For a particular combination of the LUX detector parameters in Table~\ref{table:BaseModelParametersTable}, we use our generic model (Section~\ref{sec:ModelConstruction}) to produce a set of low-energy S2 spectra, now demanding S1\(\geq\)2phd to ensure comparability to LUX S1+S2 data.\footnote{For the cathode, we also apply an S1 cutoff of 158~phd.} We produce such spectra with a range of assumed embedding fractions \(f_{e}\subset[0.0,1.0]\), and over a wide range of overall rate normalizations. A \(\chi^{2}\) minimization process is then performed over these \(f_{e}\) and rate values, comparing the generated model spectra to the LUX data selection in the region [60,1400]e-, to find a best-fit spectrum. Finally, we repeat this for all combinations of the parameters in Table~\ref{table:BaseModelParametersTable}, and in each S2 bin take the highest and lowest combinations' values as bounds of a ``best-fit model family.'' 

\begin{figure}[t!]
\centering
\includegraphics[width=\linewidth]{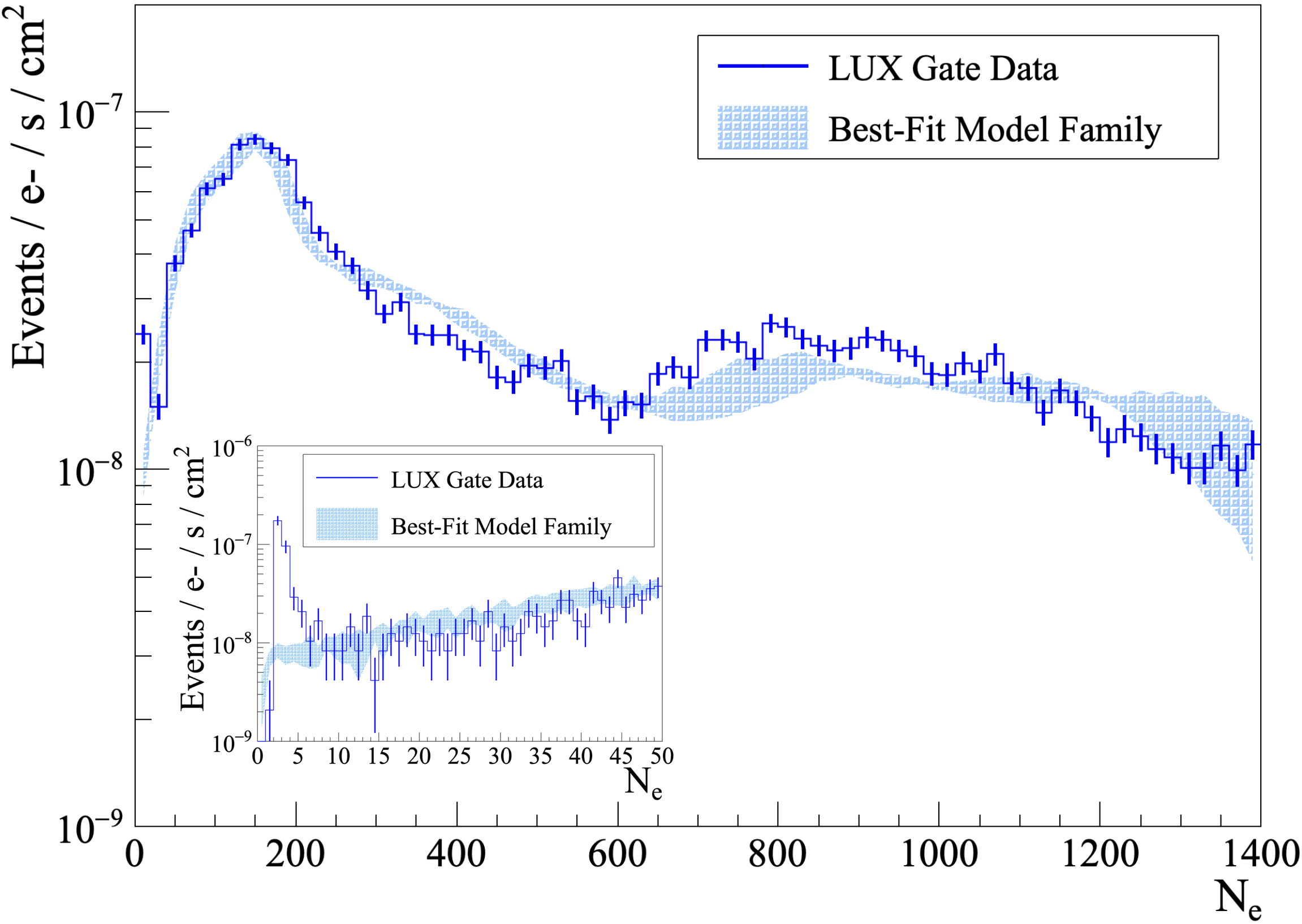}
\includegraphics[width=\linewidth]{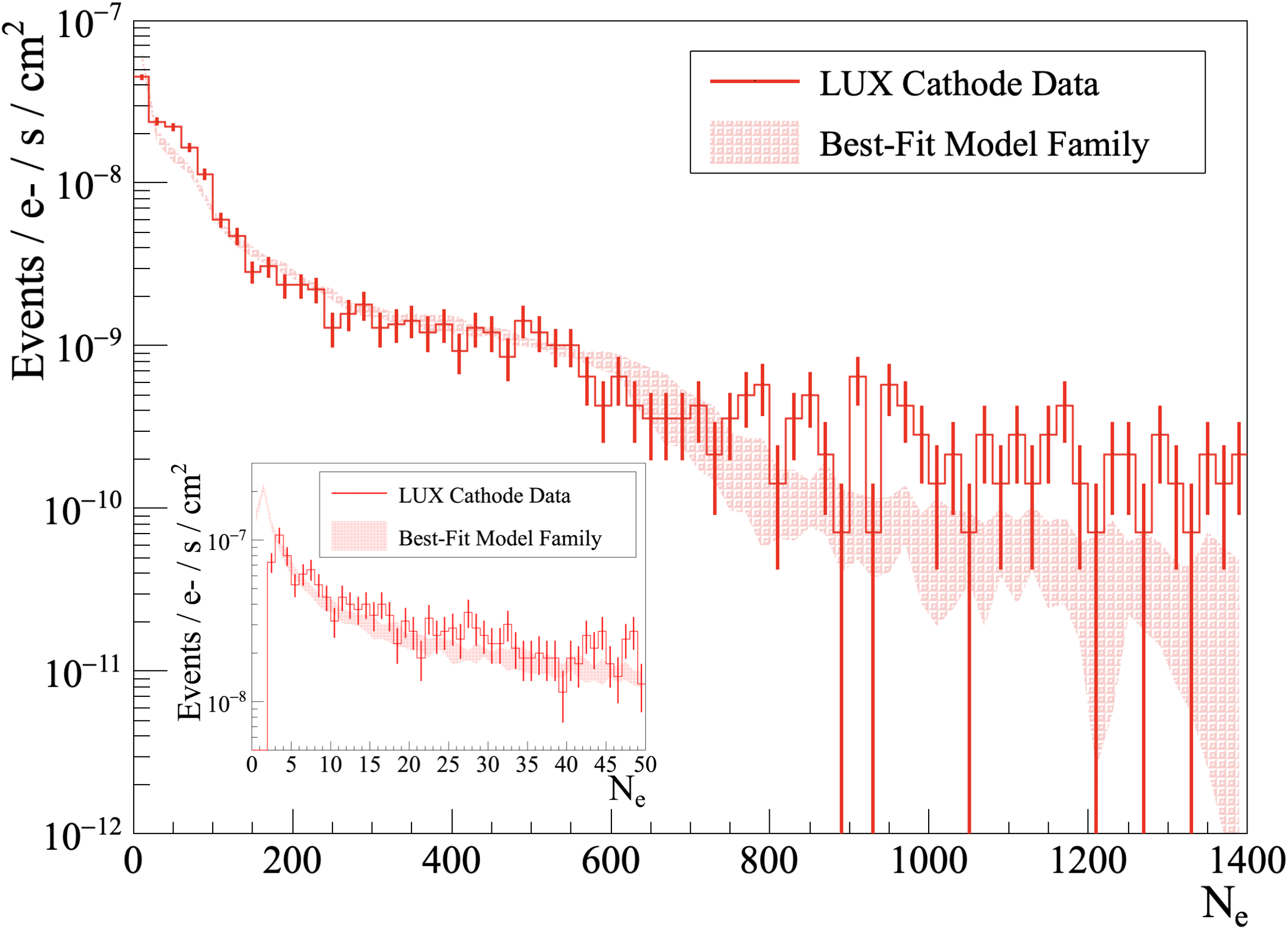}
\caption{LUX Gate (top) and cathode (bottom) measured S2 spectra (solid lines with error bars) compared to best-fit model families. The y-axis is normalized to the livetime and the total grid wire surface area, accounting for cut acceptances. See Table~\ref{table:SaturationRegionValues} for wire geometry information. Insets show the spectra in the [0,50]e- region.}
\label{fig:S1S2ValidationLUX}
\end{figure}

For the gate, the result of these fits can be seen in the top panel of Figure~\ref{fig:S1S2ValidationLUX}.  
We again observe a peak around 150~e-, which implies the presence of \(^{206}\)Pb recoils from \(^{210}\)Po decays and possibly \(^{214}\)Pb recoils from \(^{218}\)Po decays. In the S2 range of [500,1200]e-, the improved agreement between model and data (relative to LZ) is attributed in part to the absence of neutron-activated $^{125}$I, because the LUX Run3 data was acquired before DD neutron calibrations~\cite{DQThesis}. 
The model-data agreement may also appear improved relative to LZ in part because each model uses a best-fit normalization and embedding fraction for a given set of input parameters. This may cause the systematic uncertainties associated with the model to be underestimated. 
The data-model agreement remains for S2 values below 50~e-, but as with LZ, a rapid rise in few-electron backgrounds occurs in data, and is attributed to field emission from the grids accompanied by S1 production~\cite{LZ2025_EBg}.

The fits returned ranges of \(^{210}\)Pb-chain activity and embedding fractions summarized in Table~\ref{table:ResultsTable}. The radon activities on the gate are comparable to those measured in LZ. Since the total wire surface area for the LUX grids is smaller than for the LZ grids, the LUX gate therefore has higher radon contamination than the LZ gate on a per-area basis. 
Understanding the estimated embedding fraction \(f_{e}\) is a bit more subtle: the fit strategy adopted for LUX analyses requires some care in interpreting the results given that we do not explicitly include floating \(^{218}\)Po and \(^{214}\)Pb components in the fits. 
Notably, \(f_{e}\) as estimated from LUX data is lower than that obtained for LZ. This difference may be explained by 
the accumulation of emanated radon daughters on grid surfaces, with \(^{218}\)Po decays  mimicking surface \(^{210}\)Po decays in a way that could artificially deflate \(f_{e}\). 
Although the accumulation of \(^{218}\)Po and its daughters on the gate is negligible in LZ data, it may occur in LUX because the LUX liquid convection flow speed (\(\sim\)cm/s) is much larger than the electric-field-induced ion drift speed (\(\sim\)mm/s); as a result some positively charged daughters produced in the FFR through radon emanation may flow upward toward the gate rather than down toward the cathode along the FFR field lines. In addition, the cathodic nature of the gate grid and the high field strength in the extraction region will 
pull positive ions near the gate toward the grid surfaces, 
generating an ``effective grid cross section'' for positive ions larger than the grid's geometric cross section.  
Alternatively, the LUX gate could have a real surface \(^{210}\)Pb component because the LUX grids were not strung in a cleanroom environment with HEPA flow.
Taken together, these two hypotheses alleviate potential tension with a naturally near-unity embedding fraction for the \(^{210}\)Pb chain proposed at the end of Sec.~\ref{subsec:LZData}.

\begin{table*}[t!]
\centering
\setlength\extrarowheight{2pt}
\caption{Measured \(^{210}\)Pb-chain activities on the LZ and LUX cathode and gate, along with those activities normalized by the grid area. The LZ activity numbers are reproduced from Table~\ref{table:AlphaModelParametersTable} for convenience. Also shown are the estimated embedding fractions. For LZ, the best-fit embedding fractions were found to be consistently at 1.0 even with other fit parameters (like the rate normalizations for the various background components) floating within their uncertainties. For reference, estimated wire surface areas for the four grids are presented at the bottom of the table.}
\begin{tabular}{|w{c}{8.5cm}||w{c}{3.0cm}|w{c}{4.0cm}|}
\hline
Measurement & Value (LZ) & Value (LUX) \\
\hline
\hline
Total \(^{210}\)Pb activity (Gate) & 11.2 \(\pm\) 0.6 mBq & 7.1 \(\pm\) 0.2 mBq \\
Total \(^{210}\)Pb activity (Cathode) & 8.8 \(\pm\) 0.8 mBq & 2.0 \(\pm\) 0.1 mBq \\
\(^{210}\)Pb activity, normalized to wire surface area (Gate) & 7.3 \(\pm\) 0.4~\(\mu\)Bq/cm\(^{2}\) & 64.3 \(\pm\) 1.6 \(\mu\)Bq/cm\(^{2}\) \\
\(^{210}\)Pb activity, normalized to wire surface area (Cathode) & 4.3 \(\pm\) 0.4~\(\mu\)Bq/cm\(^{2}\) & 8.8 \(\pm\) 0.4 \(\mu\)Bq/cm\(^{2}\) \\
\(^{210}\)Pb chain \(f_{e}\) (Gate) & 1.0 & 0.77 \(\pm\) 0.03\\
\(^{210}\)Pb chain \(f_{e}\) (Cathode) & 1.0 & 0.0 +0.08/-0 \\
\hline
Wire surface area (Gate) & 1.53\(\times 10^{3}\)~cm\(^{2}\) & 1.10\(\times 10^{2}\)~cm\(^{2}\) \\
Wire surface area (Cathode) & 2.05\(\times 10^{3}\)~cm\(^{2}\) & 2.24\(\times 10^{2}\)~cm\(^{2}\) \\
\hline

\end{tabular}
\label{table:ResultsTable}
\end{table*}
\vspace{5mm}

For the LUX cathode, shown in the lower panel of Figure~\ref{fig:S1S2ValidationLUX}, the best fit family displays similarly good model-data consistency. The RFR losses again cause a characteristic falloff of the S2 spectrum. An ion-recoil shoulder is also present at around 60~e-, which approximately matches the corresponding shoulder in LZ when differences in detector extraction efficiency and electron lifetime are accounted for. The overall rate for the LUX cathode is much lower than that for the gate as a result of the cathode's lower radon exposure (elaborated in Sec.~\ref{subsec:LZLUXComparison}). The embedding fraction for the cathode is consistent with 0.0. 
We again ascribe this to decays of \(^{218}\)Po nuclides that have landed on the cathode surfaces and which produce the shoulder structure at 60~e- similar to surface \(^{210}\)Po decays. 
A total \(^{222}\)Rn emanation rate of 17.9~mBq was reported for LUX~\cite{BradleyPaper}, 
so even if \(O(10\%)\) of the emanated \(^{218}\)Po lands on the cathode, it would be enough to strongly contribute to the \(^{210}\)Po decay rate (2.0~mBq) extracted from the LUX cathode fit.
Similar to the gate, the LUX cathode may have a true surface component of \(^{210}\)Po incurred from production plateout in a non-cleanroom environment.

\section{Discussions}
\label{sec:Discussion}

These model-data comparisons for radon plateout backgrounds on the LZ and LUX grids have provided two primary insights worth noting for future Xe TPC experiments. First, grid backgrounds are likely dominated by \(^{210}\)Pb-chain isotopes deposited during grid production, but their spectra can include contributions from in-situ emanated radon. Second, our background model is capable of faithfully representing physics of xenon TPCs in the S1+S2 regime, giving us confidence that it may be extrapolated down into an S2-only regime to assist in low-mass dark matter searches.

\subsection{Mitigation of Grid Backgrounds}
\label{subsec:LZLUXComparison}

This modeling effort confirms that decays of \(^{210}\)Pb-chain daughters plated out during grid production are a leading source of backgrounds on the LZ and LUX grids. 
The most direct evidence for this is the \(O(10)\)~mBq rate of \(^{210}\)Po measured for the LZ grids and the \(O(1-10)\)~mBq best-fit rates of the whole \(^{210}\)Pb-chain measured for the LUX grids. Given the emanated \(^{222}\)Rn rates in LZ (\(\simeq 33\)~mBq~\cite{LZRadonFlow2025}) and LUX (17.9~mBq~\cite{BradleyPaper}), it would take at minimum several years for ``grown-in'' \(^{210}\)Pb to achieve these rates (See Figure~\ref{fig:CathodeRateEmanationScenarios} in App.~\ref{app:AdditionalModelDetails}). This production-dominated daughter deposition is consistent with observations made by other low-background experiments~\cite{OtherDrift}.

In addition, the \(^{210}\)Po alpha decay rates observed for the LZ grids (and, by secular equilibrium, the \(^{210}\)Pb rates) are within the bounds of a-priori predictions made for these grids in Ref~\cite{Linehan}. Moreover, the rates observed for the LZ gate and cathode are similar, reflecting the grids' similar air exposures during production~\cite{Linehan}. For LUX, the cathode \(^{210}\)Pb rate is substantially less than the gate rate, which is consistent with their different air exposures: the cathode was re-strung with new wire before Run3, while the gate was not. Finally,
when normalized to wire surface area, LZ grids showed overall lower plate-out rates than LUX grids, reflecting the improvements in radon control measures taken during LZ grid construction and installation, relative to those used in LUX construction~\cite{Linehan,ManninoThesis}.

\begin{figure*}[t!]
\centering
\includegraphics[width=\linewidth]{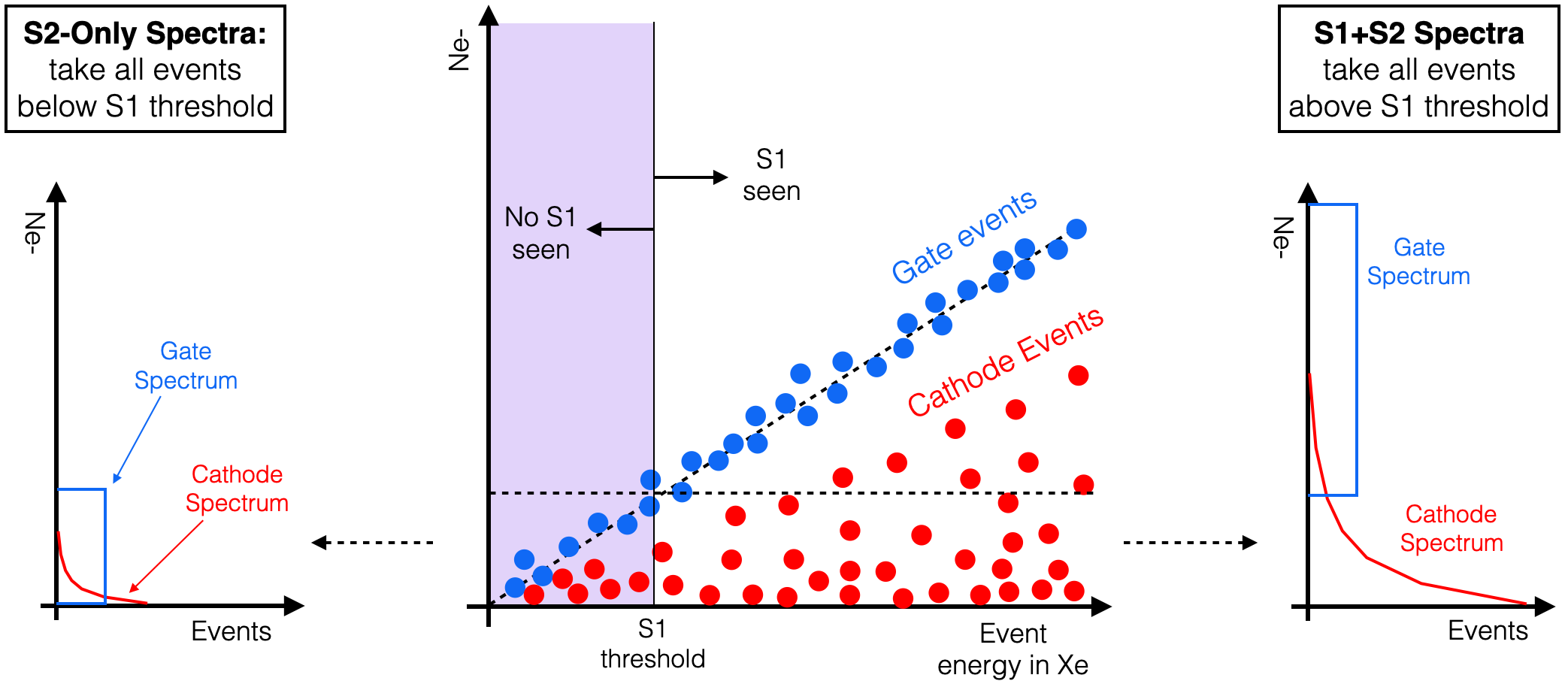}
\caption{A simplified diagram illustrating the relationship between the S2-only spectrum and the S1+S2 spectrum for gate and cathode grids running under similar electric field configurations to LUX and LZ. Red and blue dots on the center set of axes are hypothetical events from the cathode and gate, respectively. How these event populations would be placed into the S2-only spectrum or the S1+S2 spectrum based on the event S1 is shown on the rotated axes on the left and right. The S2-only spectra are filled only with events in the purple shaded region, and the S1+S2 spectra are filled only with events to the right of the purple shaded region. The ``boxy'' shapes of the gate spectra are aphysical and in reality are gentle roll-offs, but here are included for conceptual clarity. We note that the S1 threshold delineated here will correspond to a larger typical S2 than that in the TPC bulk due to the reduced light yield and light collection for events occurring near the wire surface.}
\label{fig:IntuitionDiagramForGridSpectra}
\end{figure*}

The prominence of ambient-air plateout of \(^{210}\)Pb-chain daughters highlights the importance of rigorous radon reduction measures during grid production in future xenon detectors. Such radon-reduction methods, discussed in Ref.~\cite{Linehan}, include keeping spooled wire under nitrogen purge, assembling a grid swiftly in a radon-controlled cleanroom environment, and keeping finished grids under nitrogen purge prior to and throughout integration into the TPC. As the total background rate is built from an integrated exposure to \(^{222}\)Rn after wire drawing, it is difficult to ascertain which production stage contributes most to the observed background rate. However, generally attempting to minimize the air exposure of the wires at all production stages is recommended for reducing this background.

While ambient-air plateout is likely the most critical contribution to reduce, the non-negligible impact of plateout from in-situ emanated \(^{222}\)Rn motivates additional detector design, fabrication, and operations strategies. To limit the in-situ plateout rate, it will be important to optimize materials selection for all TPC components for radiopurity to minimize the overall radon emanation, which is also important for neutron and gamma background reduction. Moreover, our efforts to use S1+S2 events to model this grid background in LZ benefited from relatively well-defined \(^{218}\)Po flow patterns in the TPC~\cite{LZRadonFlow2025}. Efforts to design dual-phase xenon TPCs that are advantageous for S2-only searches with rigorous background modeling might therefore consider minimizing turbulent flow of the xenon.

\subsection{S2-Only Grid Background}
\label{subsec:ExtendingModelToS2OnlyRegime}

The work described in Sec.~\ref{sec:S1S2Models} focuses on events with paired S1 and S2 signals for both model construction and data validation. These events can be easily rejected in a dark matter search based on their reconstructed vertical positions. 
However, the low-energy component of the grid radon backgrounds may not have observable S1s, and thus will be a dangerous background in S2-only searches for lower-mass dark matter. 
Such background S2 pulses can be indistinguishable from potential dark matter signals at the event level, but, as we will demonstrate, can be modeled for statistical subtraction from dark matter search data using the same model described in Section \ref{sec:ModelConstruction}.

By construction, the modeling process described in Section~\ref{sec:S1S2Models} produces complete S1 and S2 distributions of backgrounds from grid radioactive decays for both LUX and LZ. In this S1+S2 regime, the model generally matches the observed S2 spectra across a large energy window ([0,2400]e- in LZ and [0,1400]e- in LUX) and maintains good agreement with data in the low-S2 regime of [0,50]e- for both detectors, except for elevated few-electron rates on the gate attributed to spurious field-induced electron emission~\cite{LZ2025_EBg}. Such agreement naturally justifies extension of this model down into the S2-only regime, which is possible by using the same model output but now demanding that an event's S1\(<\)2 (3)~phd for LUX (LZ).

Much like the S1+S2 events, the S2-only spectra for the gate and cathode are expected to inherit distinct shapes from each grid's unique detector response combined with the S1 threshold, as illustrated in Figure \ref{fig:IntuitionDiagramForGridSpectra}. For gate events, one expects a roughly linear relationship between the event energy in xenon and the detected S1 and S2 sizes. This gives a somewhat flatter S2-only spectrum. 
In contrast, the cathode's RFR losses significantly weaken the correlation between an event's deposited energy and the observed S2. This weaker correlation gives the steep fall-off in the S2 spectrum, but notably also causes the shape of the S2-only spectrum to look very similar to that of the S2 spectrum from S1+S2 events, up to a scaling dependent on the choice of S1 cutoff used in the S1+S2 analysis (which sets the normalization of the S1+S2 data).

\begin{figure}[t!]
\centering
\includegraphics[width=\linewidth]{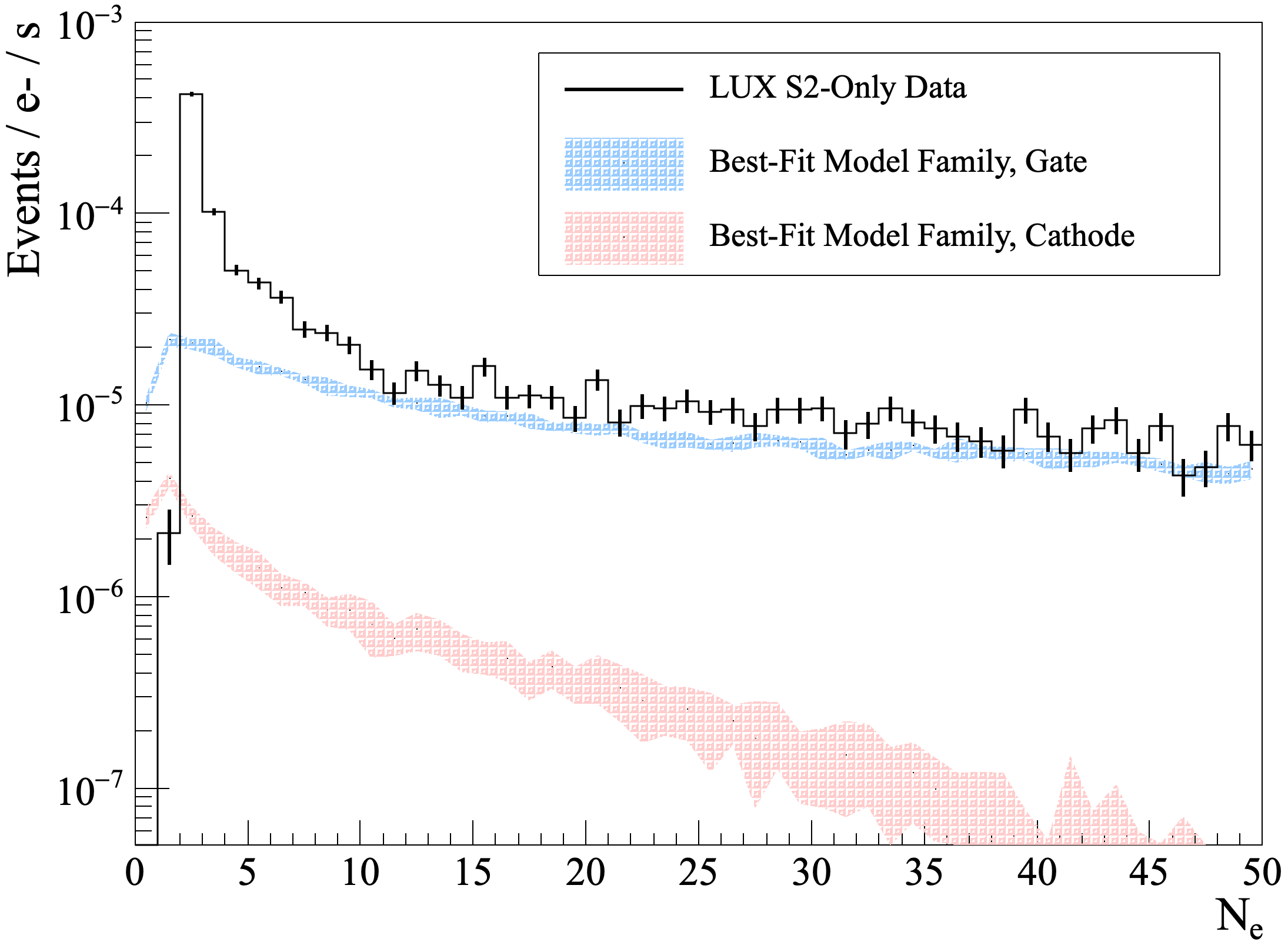}
\caption{S2-only spectra from LUX (solid line) compared to the gate and cathode models normalized using S1+S2 data. For visual clarity, the summed grid radon background contribution from both grids is not shown, but is dominated by the gate contribution.}
\label{fig:S2OnlySpectraLUX}
\end{figure}

We present estimates of this S2-only model for the LUX Run3 gate and cathode in Figure \ref{fig:S2OnlySpectraLUX}. The model family uses best-fit normalizations and embedding fractions determined from S1+S2 spectrum fits as described in Section~\ref{subsec:LUXData}. 
A separate analysis (also described in Appendix~\ref{app:SelectingLowEnergyDatasetsLUX}) was conducted on LUX Run3 data to acquire a S2-only event spectrum for comparison, which is also shown in Figure \ref{fig:S2OnlySpectraLUX}. This spectrum differs from the S2-only search data used in Ref.~\cite{KelseyLUX} in that it does not include S2 pulse-shape cuts for coarse z-fiducialization. The model tracks the data to within approximately a factor of two down to S2 values around 10e-. While we have noted that our model family likely underestimates systematic uncertainties due to the relatively simple model used for LUX, this general agreement nonetheless makes explicit the role of grid backgrounds in defining the shape of the S2-only spectrum down to 10e-. 
The sharp rise in the spectrum below 10e- could be explained as backgrounds other than grid radon-chain decays. Here, as in the S1+S2 case, we propose that field-induced electron emission with near-wire multiplication begins to play a larger role in shaping the S2-only spectrum~\cite{BaileyThesis,LZ2025_EBg}, 
and in the $<$3 e- region pile-up of delayed electrons could become dominant~\cite{LUXEBackgrounds,LZ2025_EBg}. 

The rise in the spectrum in Figure~\ref{fig:S2OnlySpectraLUX} as S2 falls from 50e- down to 10e- is partially attributed to the presence of surface roughness (``teeth'') in the model geometry. Loss of particle energy to surface roughness has also been used to describe rises in low-energy event spectra in other low-background dark matter experiments~\cite{CRESST,CRESSTRoughness}, and at next-to-leading order may also include effects from Q/L yield variations or electron reabsorption in field-shadowed pockets created by such roughness. 
However, several variations in roughness dimension were attempted without successful reproduction of the remaining excess below 10e-. Although the next-to-leading order effects are not explored in this work, the fixed ``budget'' of grid radiogenic decays also implies that the model-data agreement below 10- could only be improved at the cost of a degradation of data-model agreement at higher S2 values. These insights provide further evidence that other backgrounds may dominate here. As this sub-10e- region is the one from which S2-only searches for low-mass dark matter derive much of their sensitivity, such searches will benefit from further exploring the spatial and temporal correlation of field-induced grid electron emission, as well as the significant probability for these electrons to be accompanied by photon production, as studied in \cite{LZ2025_EBg}.

Ultimately, we find that this gate and cathode radon plateout background provides a large fraction of the S2-only background expected in dual-phase TPCs, though it may be insufficient to explain all observed event rate down to the lowest S2s. Given radon backgrounds' dominance here, S2-only backgrounds may be significantly reduced through cuts on S2 pulse shape that provide coarse \textit{z}-fiducialization and preferentially exclude cathode and gate events over bulk events. Such cuts have already been demonstrated in LUX and other experiments~\cite{KelseyLUX,IonizationOnlyPandaX4T_2023,wang2025searchlightinelasticdark,XENON1TS2O}. These cuts, along with the work in this paper and further model-building needed in the region below 10e-~\cite{LZ2025_EBg}, will be instrumental in boosting the sensitivity and discovery potential of S2-only searches in future experiments.

\section{Conclusions}
\label{sec:Conclusions}

In this work we have performed a detailed study of radon backgrounds on high voltage electrodes in dual-phase xenon TPCs. We have developed an approximate first-principles model of the S2 spectra that are expected from these backgrounds. In the S1+S2 regime, we observe that our model largely matches the structure in gate and cathode S2 spectra measured by both the LZ experiment (WS2022) and the LUX experiment (Run3). This analysis also provides evidence that these backgrounds are likely dominated by plateout of \(^{210}\)Pb during grid production and installation, and reinforces the need to limit radon exposure during production of future experiments' electrodes. Fits of the LZ and LUX data further indicate that events from plateout of in-situ-emanated \(^{222}\)Rn, though subdominant, also play a role in shaping the S2 backgrounds from these grids.

We have also compared this model with S2-only data from the LUX experiment's Run3. We observe that the model reasonably tracks the data down to S2 values of approximately 10e-. We conclude that grid radon backgrounds define the shape of the S2-only spectrum in an S2 range above about 10e-, though there are a number of systematic uncertainties that make constructing a precise model spectrum shape and normalization challenging. Below 10e-, other backgrounds such as field induced emission and single electron pileup can dominate over these radon decays. Future analysis of the S2-only spectrum in LZ will refine our understanding of both radon-related and other S2-only backgrounds. Even though further work is needed to build a comprehensive understanding of all S2-only backgrounds, we conclude that grid radon decays nonetheless form an important foundation for a general background model for S2-only analyses in dual-phase TPCs.

\begin{acknowledgments}

The research supporting the LZ work took place in part at the Sanford Underground Research Facility (SURF) in Lead, South Dakota. Funding for this work is supported by the U.S. Department of Energy, Office of Science, Office of High Energy Physics under Contract Numbers DE-AC02-05CH11231, DE-SC0020216, DE-SC0012704, DE-SC0010010, DE-AC02-07CH11359, DE-SC0015910, DE-SC0014223, DE-SC0010813, DE-SC0009999, DE-NA0003180, DE-SC0011702, DE-SC0010072, DE-SC0006605, DE-SC0008475, DE-SC0019193, DE-FG02-10ER46709, UW PRJ82AJ, DE-SC0013542, DE-AC02-76SF00515, DE-SC0018982, DE-SC0019066, DE-SC0015535, DE-SC0019319, DE-SC0025629, DE-SC0024114, DE-AC52-07NA27344, \& DE-SC0012447; the U.S. National Science Foundation under Grants No. PHY-0750671, PHY-0801536, PHY-1003660, PHY-1004661, PHY-1102470, PHY-1312561, PHY-1347449, PHY-1505868, and PHY-1636738; the Research Corporation Grant No. RA0350; the Center for Ultra-low Background Experiments in the Dakotas (CUBED); and the South Dakota School of Mines and Technology (SDSMT). This research was also supported by the UKRI’s Science \& Technology Facilities Council under award numbers ST/W000490/1, ST/W000482/1, ST/W000636/1, ST/W000466/1, ST/W000628/1, ST/W000555/1, ST/W000547/1, ST/W00058X/1, ST/X508263/1, ST/V506862/1, ST/X508561/1, ST/V507040/1, ST/W507787/1, ST/R003181/1, ST/R003181/2,  ST/W507957/1, ST/X005984/1, ST/X006050/1; Portuguese Foundation for Science and Technology (FCT) under award numbers PTDC/FIS-NUC/1525/2014 and PTDC/FIS-PAR/2831/2020; the Institute for Basic Science, Korea (budget number IBS-R016-D1); the Swiss National Science Foundation (SNSF)  under award number 10001549. This research was supported by the Australian Government through the Australian Research Council Centre of Excellence for Dark Matter Particle Physics under award number CE200100008. We acknowledge additional support from the UK Science \& Technology Facilities Council (STFC) for PhD studentships and the STFC Boulby Underground Laboratory in the U.K., the GridPP~\cite{faulkner2006gridpp,britton2009gridpp} and IRIS Collaborations, in particular at Imperial College London and additional support by the University College London (UCL) Cosmoparticle Initiative, and the University of Zurich. We acknowledge additional support from the Center for the Fundamental Physics of the Universe, Brown University. K.T. Lesko acknowledges the support of Brasenose College and Oxford University. This research used resources of the National Energy Research Scientific Computing Center, a DOE Office of Science User Facility supported by the Office of Science of the U.S. Department of Energy under Contract No. DE-AC02-05CH11231. We gratefully acknowledge support from GitLab through its GitLab for Education Program. The University of Edinburgh is a charitable body, registered in Scotland, with the registration number SC005336. The assistance of SURF and its personnel in providing physical access and general logistical and technical support is acknowledged. We acknowledge the South Dakota Governor's office, the South Dakota Community Foundation, the South Dakota State University Foundation, and the University of South Dakota Foundation for use of xenon. We also acknowledge the University of Alabama for providing xenon. For the purpose of open access, the authors have applied a Creative Commons Attribution (CC BY) license to any Author Accepted Manuscript version arising from this submission. Finally, we respectfully acknowledge that we are on the traditional land of Indigenous American peoples and honor their rich cultural heritage and enduring contributions. Their deep connection to this land and their resilience and wisdom continue to inspire and enrich our community. We commit to learning from and supporting their effort as original stewards of this land and to preserve their cultures and rights for a more inclusive and sustainable future. 

The LUX work was partially supported by the U.S. Department of Energy (DOE) under Award No. DE-AC02-05CH11231, DE-AC05-06OR23100, DE-AC52-07NA27344, DE-FG01-91ER40618, DE-FG02-08ER41549, DE-FG02-11ER41738, DE-FG02-91ER40674, DE-FG02-91ER40688, DE-FG02-95ER40917, DE-NA0000979, DE-SC0006605, DE-SC0010010, DE-SC0015535, and DE-SC0019066; the U.S. National Science Foundation under Grants No. PHY-0750671, PHY-0801536, PHY-1003660, PHY-1004661, PHY-1102470, PHY-1312561, PHY-1347449, PHY-1505868, and PHY-1636738; the Research Corporation Grant No. RA0350; the Center for Ultra-low Background Experiments in the Dakotas (CUBED); and the South Dakota School of Mines and Technology (SDSMT).
Laborat\'{o}rio de Instrumenta\c{c}\~{a}o e F\'{i}sica Experimental de Part\'{i}culas (LIP)-Coimbra acknowledges funding from Funda\c{c}\~{a}o para a Ci\^{e}ncia e a Tecnologia (FCT) through the Project-Grant PTDC/FIS-NUC/1525/2014. Imperial College and Brown University thank the UK Royal Society for travel funds under the International Exchange Scheme (IE120804). The UK groups acknowledge institutional support from Imperial College London, University College London, the University of Sheffield, and Edinburgh University, and from the Science \& Technology Facilities Council for PhD studentships R504737 (EL), M126369B (NM), P006795 (AN), T93036D (RT) and N50449X (UU). This work was partially enabled by the University College London (UCL) Cosmoparticle Initiative. The University of Edinburgh is a charitable body, registered in Scotland, with Registration No. SC005336.
This research was conducted using computational resources and services at the Center for Computation and Visualization, Brown University, and also the Yale Science Research Software Core.
We gratefully acknowledge the logistical and technical support and the access to laboratory infrastructure provided to us by SURF and its personnel at Lead, South Dakota. SURF was developed by the South Dakota Science and Technology Authority, with an important philanthropic donation from T. Denny Sanford. SURF is a federally sponsored research facility under Award Number DE-SC0020216.

\end{acknowledgments}

\appendix
\section{Appendix: Additional Model Details}
\label{app:AdditionalModelDetails}

Four additional modeling details are worth further discussion: our decision to not model the anode and shield grids, our decision not to model an additional rate of \(^{210}\)Pb decays from in-situ plate-out, the assumed azimuthal location of the various daughters on our wires, and the wire surface roughness used in our BACCARAT simulations.

In this work we have focused primarily on modeling the gate and cathode backgrounds, under the claim that these two grids are the only ones that will contribute meaningfully to S2-only backgrounds. Decays from the bottom shield grid will contribute events with only an S1 signal, as the grid is anodic and freed charges will drift back onto it rather than drifting up to the gas xenon layer. Much like the cathode, the \(^{210}\)Po rates should be measurable from the S1 spectrum and provide an estimate of the \(^{210}\)Pb chain. While these decays will not contribute S2s in any appreciable amount, they will contribute to the detector's S1-only spectrum. While the low-energy \(^{210}\)Pb decays may still pose an issue for low-energy S1-S2 analyses via their ability to form accidental-coincidence backgrounds~\cite{LZFirstResults}, we nevertheless do not model them here. Decays from the anode (and in the case of LUX, the top shield grid) will produce S1s and electroluminescence, but due to the low gas density track lengths are significantly longer in the gas than they are in the liquid. This, combined with the nontrivial field configuration in the near-anode region, may create long, distorted S2 pulse shapes that are prohibitively challenging to model from first principles without significant computational expense. However, since these pathological S2 pulse shapes can be easily cut from an S2-only search, we also do not model these backgrounds.

\begin{figure}[t]
\centering
\includegraphics[width=\linewidth]{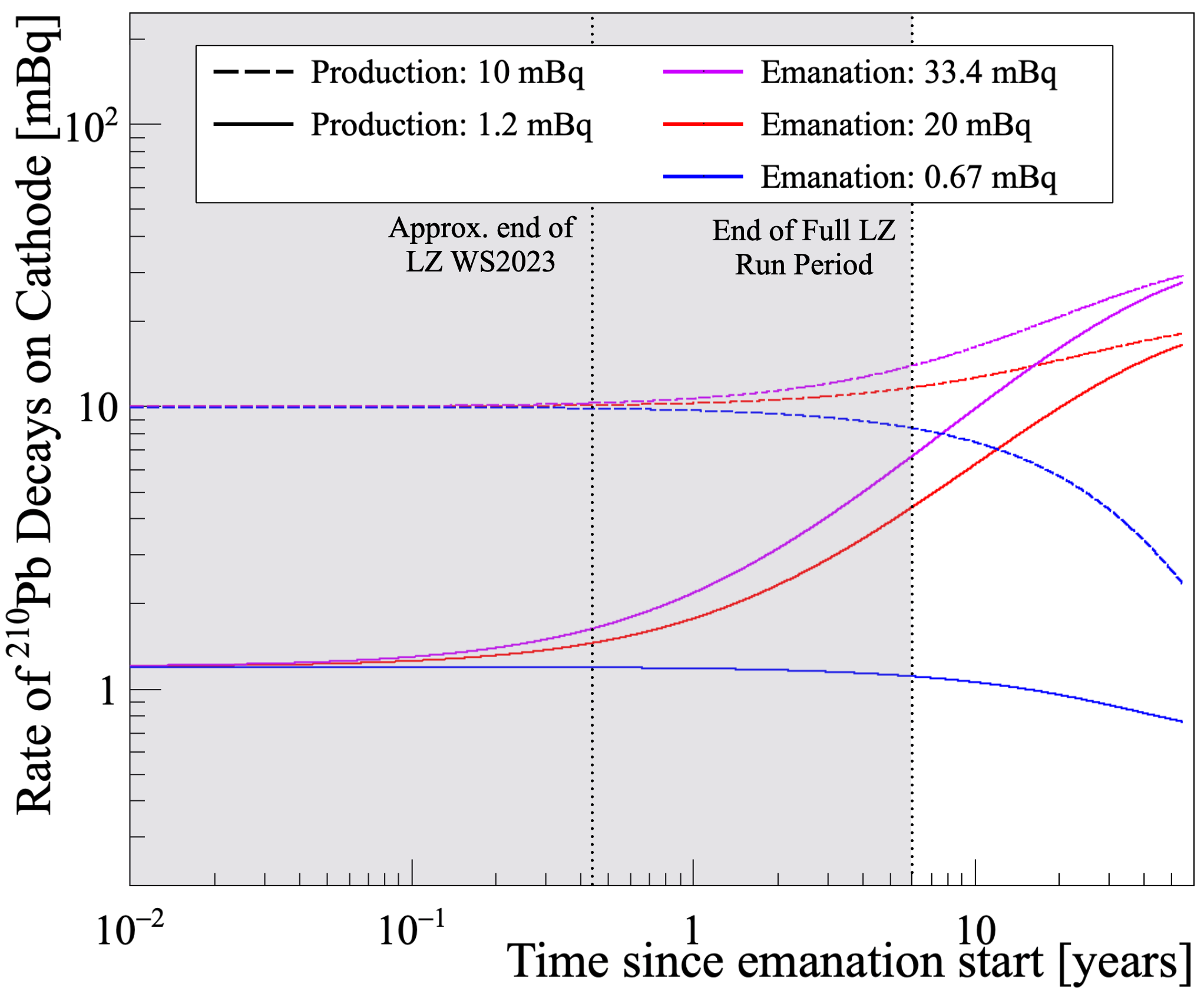}
\caption{Estimated rates of \(^{210}\)Pb chain decays for scenarios with various grid-production-deposited daughters and emanation-deposited daughters~\cite{LinehanThesis}. Solid lines show scenarios in which the overall activity contributed from plate-out during production is ``low'' -- the value of 1.2~mBq is taken from the optimistic cathode plate-out scenario in Ref~\cite{Linehan}. Dashed lines are the same set of scenarios in which that production plate-out is about a factor of 10 higher, closer to the upper bound in Ref~\cite{Linehan} and the \(\mathcal{O}\)(10~mBq) rates measured in this work (Table~\ref{table:AlphaModelParametersTable}). The ``emanation'' value is the total rate of \(^{210}\)Pb deposition on the cathode from emanated \(^{222}\)Rn, which cannot be higher than the overall emanated \(^{222}\)Rn rate itself. In this, we conservatively assume that all emanated \(^{222}\)Rn atoms deposit a \(^{210}\)Pb ion onto the cathode.}
\label{fig:CathodeRateEmanationScenarios}
\end{figure}

In this work we have also referenced a decision not to model an additional \(^{210}\)Pb activity due to in-situ plate-out from \(^{222}\)Rn emanated from within the cryostat. This is justified due to the 22.2-year \(^{210}\)Pb lifetime, which is long compared to typical experimental schedules. This implies that unless the rate of \(^{210}\)Pb from plate-out during production is highly subdominant to the \(^{222}\)Rn emanation rate, it will take several years for this emanated radon to produce a meaningful relative change in the \(^{210}\)Pb rate (Figure~\ref{fig:CathodeRateEmanationScenarios}). In Figure~\ref{fig:CathodeRateEmanationScenarios}, we assume that the ``turn-on'' of plateout on the cathode is when the standard TPC operating fields are established. This is an approximation -- realistically, some \(^{222}\)Rn emanation and on-grid plateout will have occurred due to purely diffusion-related transport in the approximately two year period between installation and in-liquid electric field commissioning. However, we estimate that this would have been subdominant to the plateout that occurred during production and also during normal TPC operation, where electric fields preferentially funnel daughters onto the wires. From another angle, Figure 9 shows that for warm radon emanation rates reasonably close to those observed during normal operation~\cite{ChottLZEmanation}, it would take 9 years for such emanation to grow-in a \(^{210}\)Pb cathode rate matching the \(\mathcal{O}\)(10~mBq) rate measured with the alpha analysis in this work (Sec.~\ref{subsec:LZData}, Table~\ref{table:AlphaModelParametersTable}). This observation also underlies our justification that the measured, \(O\)(10~mBq) rates of \(^{210}\)Po in LZ are in fact primarily from production plate-out: it would take far longer than the current run-time of LZ to produce these rates via emanation alone, given measured emanation rates.

Our model assumes a few different conditions for the distribution of the plated-out radon daughters around the wire. For \(^{210}\)Pb-chain daughters from ambient-air plate-out, the exact distribution is challenging to know \textit{a-priori}. However, due to the relative stillness of ambient air, one might reasonably expect this distribution to be azimuthally symmetric. Since there is also a weak downward flow of HEPA-filtered air through the wires from the cleanroom production environment~\cite{Linehan}, we add a slight asymmetry to this as discussed in the text. For the set of daughters from in-situ plate-out in a TPC dominated by electrostatic drift (and not convection), the distribution is easier to intuit: charged daughters will drift along field lines until their termination point. Since the field strength of the forward field region (FFR) is about 10\(\%\) that in the reverse field region, only about the top 10\(\%\) of the wire is expected to be the termination point for the field lines carrying daughters from the FFR, with the other 90\(\%\) being termination points for field lines from the RFR. This justifies our approximation that the top 10\(\%\) of the cathode wires inherit the bulk of the downward drifting daughters in LZ.\footnote{If we were to have used the same analysis technique for LUX, then this assumption would not hold, as the convective processes in LUX were dominant over electrostatic transport, unlike in LZ.}

\begin{figure}[t]
\centering
\includegraphics[width=\linewidth]{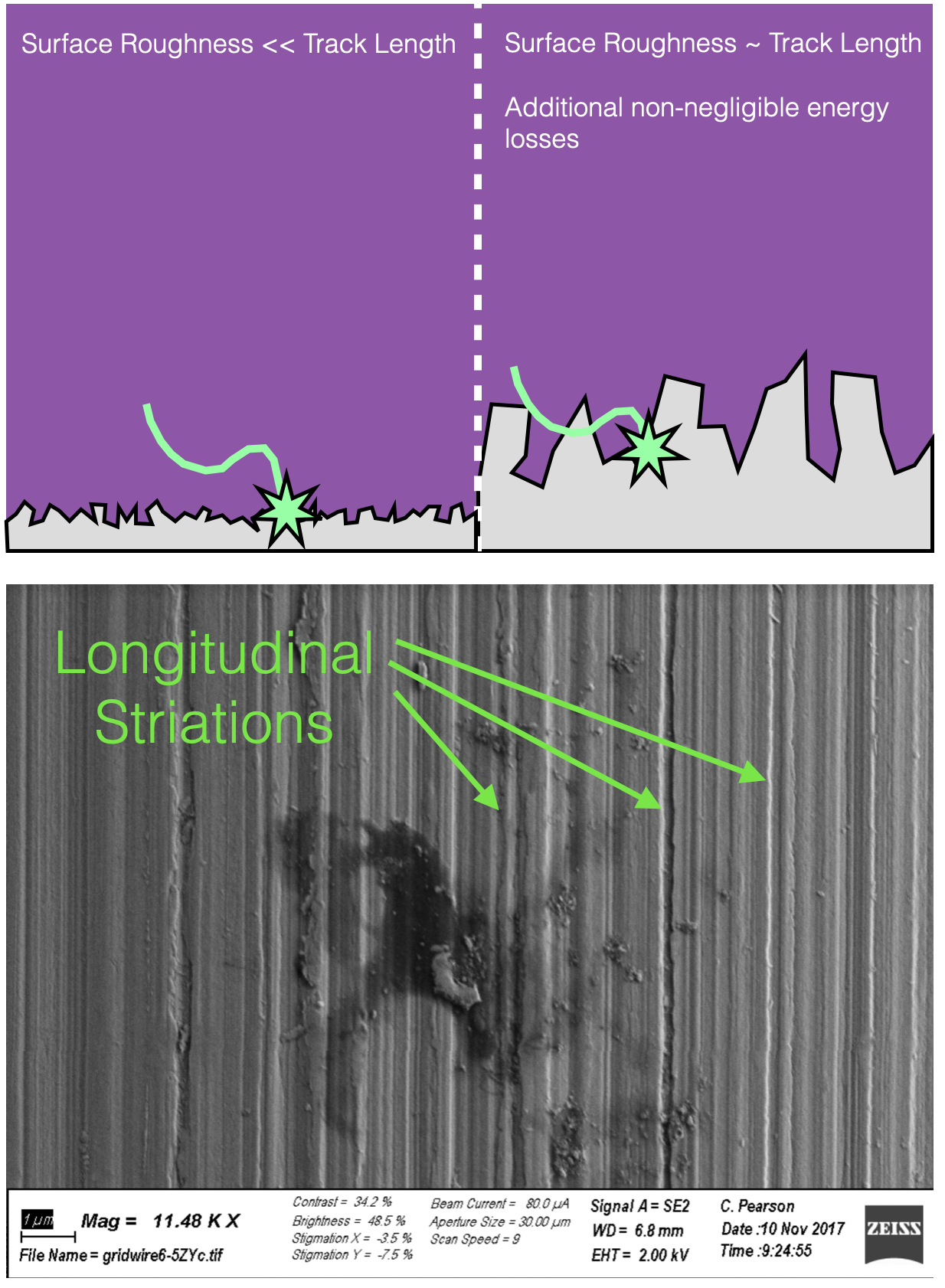}
\caption{Top: a diagram of how a smoother surface (left) causes minimal in-wire energy loss for a low-energy track, while a rougher surface may cause more substantial in-wire energy loss. Bottom: an SEM image of a sample of wire used in prototype grid development. This wire is the same type as and is from the same vendor as the wire used in LZ’s final grids. The scale is shown in the bottom left of the image. Imperfections like the central smudge here were common on SEM images of wires that were taken on test samples during initial grid prototyping, but the exact nature of such imperfections (i.e. oxide layers, contamination, etc.) and the prevalence of them on the final wires used in LZ was not thoroughly explored.}
\label{fig:SurfaceRoughnessPlusSEM}
\end{figure}

Finally, our assumptions of surface roughness are worth discussing. Modeling this surface roughness enables a proper accounting of the lowest-energy events, since small variations in the amount of material in the path of, say, a low-energy electron or ion recoil can degrade the measured S1 and S2 signals and skew events to lower S2 values. This is particularly important for capturing subtle effects of the wire surface on the subset of \(^{222}\)Rn-chain decays with \(\mathcal{O}(\mu m)\) characteristic length scales that are not much larger than characteristic wire surface roughness length scales (Table~\ref{table:RelevantDecayProducts}). When track lengths of decay products are on the same scale as the characteristic roughness of the material, substantial fractions of the energy deposited may be lost in the SS wire (Figure \ref{fig:SurfaceRoughnessPlusSEM}, top). In addition to energy losses, surface roughness of the wires may change the light and charge yields from near-wire energy depositions in the xenon. In concave ``troughs'' caused by wire roughness, field shadowing may reduce the overall field in that region, increasing (decreasing) the light (charge) yield relative to high-points on the wire, but such troughs may also force light and electrons to scatter additional times into the wire, increasing the re-absorption probability and potentially reducing both charge and light yield. Moreover, local field nonuniformities from surface roughness may also lead to preferential in-situ plateout of charged daughters on the tops of the teeth where the field is higher, inducing additional next-to-leading order effects in the S2 spectrum from these decays. While these effects are not explicitly handled in this work, they may be systematics that need future exploration if improvements to this current model are desired.

To attempt to qualitatively charaterize this surface roughness effect in LZ, we may look at SEM images like that in the bottom of Figure \ref{fig:SurfaceRoughnessPlusSEM}, which gives an approximate lateral scale for surface roughness. If one assumes the vertical (radial) roughness is comparable to the lateral variations, one finds for LZ's wires roughness on the scale of 100~nm. As betas around 1~keV give track lengths in liquid xenon roughly at the 100~nm scale~\cite{ESTAR,Dahl}, this roughness will begin to non-negligibly skew the S2 spectrum to lower S2s below a few keV in decay energy. Additional tooth profiles of 20~nm x~20~nm and 100~nm x 300~nm (deeper than wide) were attempted, and though they marginally modified the energy spectrum in the few-electron regime, the differences in spectrum from varying tooth profile were less prominent than the difference between any tooth model and the smooth-wire model. These roughness-induced losses are suspected culprits of low-energy excesses in existing dark matter experiments with a variety of detector architectures~\cite{CRESST,CRESSTRoughness}, indicating the importance of accounting for them in a low-energy background model. 

Despite the importance of modeling these, doing so rigorously is challenging in practice. The above assumption that the vertical scale of roughness matches the lateral scale of roughness is arbitrary, and without an atomic-force measurement (AFM) the three-dimensional roughness cannot be fully characterized. Moreover, due to the impracticality of accurately measuring the surface roughness of wires on the meter-to-kilometer scale required for the current generation and future generations of TPCs, this surface roughness is a somewhat irreducible systematic uncertainty in any S2-only background model for grid radiogenic backgrounds. 

\section{Appendix- LZ Data Selection}
\label{app:SelectingLowEnergyDatasetsLZ}

Low-energy gate and cathode S2 spectra are extracted from 79.6 days of the LZ WS2022 data, a subset of the complete 89 day exposure. We first acquire clean datasets of low-energy gate and cathode events. The following cuts are used to select a clean sample of grid events:
\begin{enumerate}
\item \textbf{Single scatter selection:} only events that are consistent with single scatters are chosen, using the same cut described in Ref~\cite{LZSR1}.
\item \textbf{Data quality cuts and vetoes:} A fiducial cut in S2 \textit{xy} space is used to remove peripheral events with \(R_{S2}>68\)~cm. A more conservative fiducial cut excluding events with \(R_{S2}>55\)~cm from the wall was also applied in the gate analysis to remove pathological ``glue ring'' events between the gate and anode rings that are systematically misreconstructed into the TPC volume due to optical shadowing. Three time-based exclusion cuts are used to remove periods of high activity after muons and large S2s and during grid electron emission flare-ups. 
\item \textbf{Targeted exclusions:} Scatters in the gas above the anode grid often mimicked events from daughters on the gate, and are cut by removing events where the S1 or S2 had an anomalously high pulse top-bottom-asymmetry (TBA). A separate conservative cut in log(S1), log(S2) space is also used to target gate-lookalike events produced by large S1s photoionizing the gate wires. A few other timing- and S1-shape-based cuts are also applied to remove events with pathological S1s formed by misreconstruction of other pulses.
\item \textbf{S1 Cutoff:} As for simulated cathode events, cathode events with \(S1>158\)~phd are cut.
\item \textbf{S2 ROI}: The S2 region of interest for  the gate and cathode analyses is [0,2400]e-. For the gate analysis, this corresponds to an energy range up to approximately 56~keV electron-equivalent (keV\(_{ee}\)). While this range is again more conservative than the 150~keV\(_{ee}\) model energy cutoff, it is chosen to further restrict analysis to an energy range where environmental gammas coming from the gas are subdominant to the grid radiogenic background. 
\item \textbf{Grid selection}: A cut in an event's drift time is used to perform the final selection of gate or cathode events. Events with drift time between [1.95,3.10] \(\mu\)s are kept for the gate and events with drift time between [949.5,954] \(\mu\)s are kept for the cathode. 
\end{enumerate}
The main cut acceptances worth noting are livetime losses from vetoes and time-based exclusions (which together reduce the 79.6 days of data to a final 54.1 livedays), XY fiducial cut acceptance (58\(\%\) for gate, 86\(\%\) for cathode), and the grid selection cut acceptance, which is about 94.9\(\%\) for the gate and effectively 100\(\%\) for the cathode. Non-grid background contamination in these spectra are estimated at the sub-1\(\%\) level for the cathode and 8\(\%\) level for the gate, where there is a high rate of gammas entering the liquid from the gas. The ``LZ Data'' lines in Figures~\ref{fig:GateModelBreakdownPlusFamily} and~\ref{fig:CathodeModelBreakdownPlusFamily} show the resulting spectra after these cuts.\footnote{For the S1+S2 event spectra in this paper, we apply cut acceptances, including livetime exclusions, to the model and compare to the data, rather than correcting the data.}

\section{Appendix- LUX Data Selection}
\label{app:SelectingLowEnergyDatasetsLUX}

Low-energy gate and cathode S2 spectra, as well as S2-only data, are extracted from a subset of LUX Run3 data. Selection cuts, many of which are similar to those used in the LZ analysis, are listed below.
\begin{enumerate}
\item \textbf{S1 Requirement:} For the S1+S2 analysis, events require a visible S1 pulse. An S1 is defined as a pulse with signal in two PMT channels and a total S1 area of at least 1~phd.
\item \textbf{Data quality cuts and vetoes:} an \textit{xy} fiducial cut removes events with R\(_{S2}>\)16~cm. Time-based vetoes are used to remove periods of sporadic high-rate electron emission from grids as well as periods of high photon and electron emission after large S2s. A time-based veto is also used to remove blocks of data during which \(^{83m}\)Kr calibrations were actively being performed in the detector.
\item \textbf{Targeted population exclusions:} scatters in the gas and other pathological events are removed using a cut on the goodness-of-fit parameter associated with the S2 \textit{xy} position reconstruction, as well as a cut on the relative areas of the S1/S2 pulses and all other pulses in the event.
\item \textbf{S1 Cutoffs:} An S1 cutoff of 158~phd is used for the cathode to keep the events populating the cathode spectrum below approximately 80~keV\(_{ee}\), similar to LZ.
\item \textbf{S2 ROI:} An S2 ROI of [0,1400]e- is used to limit contamintaion from external gamma backgrounds. This is less broad than the LZ ROI due to the different extraction efficiency and drift losses, but represents approximately the same energy range.
\item \textbf{Grid Selection Cuts:} gate decays are selected by keeping events with drift times of [0,7]~\(\mu\)s. Cathode decays are selected by keeping events with drift times \(>321~\mu\)s.
\end{enumerate}

The time-based cuts act as a loss in livetime, resulting in an overall livetime of 54 days. The acceptance of the \textit{xy} fiducial cut was 44$\%$ for gate events, but 64$\%$ for cathode events due to a slight nonuniformity of the field lines in LUX that pushes the detector wall to lower S2 radius with decreasing \textit{z}-position. The S2 quality cuts are expected to cut a negligible fraction of gate and cathode events. Figure \ref{fig:S1S2ValidationLUX} shows the results of these selection cuts.

The analysis cuts used to define the S2-only dataset shown in Figure~\ref{fig:S2OnlySpectraLUX} are identical to those described above, with the exception that they require no visible S1 pulse and exclude the S1 cutoff and the drift time cuts used for the gate and cathode selections.

\bibliographystyle{apsrev}
\bibliography{main}

\end{document}